\documentclass[prd,amssymb,amsmath,amsfonts,nofootinbib,reprint,longbibliography,superscriptaddress]{revtex4-1}

\usepackage{graphicx}
\usepackage{lmodern}
\usepackage{amsmath,amssymb}
\usepackage{mathrsfs}
\usepackage{amsfonts}
\usepackage[utf8]{inputenc}
\usepackage{url}
\usepackage[colorlinks]{hyperref}
\usepackage[dvipsnames]{xcolor}
\usepackage{multirow}
\usepackage[normalem]{ulem}
\usepackage{float}
\usepackage{marvosym}
\usepackage{enumerate}
\usepackage{color,soul}
\usepackage{placeins}
\usepackage{bm}
\usepackage{color}
\usepackage{commath}
\allowdisplaybreaks
\usepackage{multirow}
\usepackage{cleveref}

\def\TEOBResumS{\texttt{TEOBResumS}}
\def\TEOBResumSGIOTTO{\texttt{TEOBResumS-GIOTTO}}
\def\TEOBResumSecce{\texttt{TEOBResumS-Dal\'i}}
\def\bajes{\texttt{bajes}}
\def\Boldtheta{\boldsymbol{\theta}}
\def\Boldd{\textbf{d}}

\definecolor{cyan}{rgb}{0,0.9,0.9}
\definecolor{orange}{rgb}{0.9,0.5,0}
\definecolor{magenta}{rgb}{1,0,1}
\definecolor{purple}{rgb}{0.8,0.4,0.8}
\definecolor{gray}{rgb}{0.8242,0.8242,0.8242}
\definecolor{dodgerblue}{rgb}{0.12, 0.56, 1.0}

\newcommand{\bham}{\affiliation{School of Physics and Astronomy and Institute for Gravitational Wave Astronomy, University of Birmingham, Edgbaston, Birmingham, B15 2TT, United Kingdom}}
\newcommand{\infn}{\affiliation{INFN Sezione di Torino, Via P. Giuria 1, 10125 Torino, Italy}}
\newcommand{\turin}{\affiliation{Dipartimento di Fisica, Universit{\`a} di Torino, Via P. Giuria 1, 10125 Torino, Italy}}
\newcommand{\jena}{\affiliation{Theoretisch-Physikalisches Institut, Friedrich-Schiller-Universit{\"a}t Jena, 07743, Jena, Germany}}
\newcommand{\ihes}{\affiliation{Institut des Hautes Etudes Scientifiques, 91440 Bures-sur-Yvette, France}}

\hypersetup{
    colorlinks=true,
    linkcolor=dodgerblue, 
    urlcolor =dodgerblue}

\begin{document}

\title{Inferring eccentricity evolution from observations of coalescing binary black holes}

\author{Alice Bonino}
\bham
\author{Rossella Gamba}
\jena
\author{Patricia Schmidt}
\bham
\author{Alessandro Nagar}
\infn
\ihes
\author{Geraint Pratten}
\bham
\author{Matteo Breschi}
\jena
\author{Piero Rettegno}
\bham
\infn
\turin
\author{Sebastiano Bernuzzi}
\jena

\begin{abstract}
The origin and formation of stellar-mass binary black holes remains an open question that can be addressed by precise measurements of the binary and orbital parameters from their gravitational-wave signal. Such binaries are expected to circularize due to the emission of gravitational waves as they approach merger. However, depending on their formation channel, some binaries could retain a non-negligible eccentricity when entering the frequency band of current gravitational-wave detectors, which will decay as the binary inspirals. 
In order to meaningfully measure the eccentricity in an observed gravitational-wave signal, two main ingredients are then necessary: an accurate waveform model that describes binaries on eccentric orbits, and an estimator to measure the non-circularity of the orbit as a function of frequency. 
In this work we first demonstrate the efficacy of the improved \TEOBResumS{} waveform model for eccentric coalescing binaries with aligned spins. We validate the model against mock signals of aligned-spin binary black hole mergers and quantify the impact of eccentricity on the estimation of other intrinsic binary parameters. We then perform a fully Bayesian reanalysis of GW150914 with the eccentric waveform model. We find (i) that the model is reliable for aligned-spin binary black holes and (ii) that GW150914 is consistent with a non-eccentric merger although we cannot rule out small values of initial eccentricity at a reference frequency of $20$ Hz. Secondly, we present a systematic, model-agnostic method to measure the orbital eccentricity and its evolution directly from the gravitational-wave posterior samples. This method mitigates
against the contamination of eccentricity measurements through the use of gauge-dependent quantities and has the advantage of allowing for the direct comparison between different analyses, as the definition of eccentricity may differ between models. 
Our scheme can be applied even in the case of small eccentricities and can be adopted straightforwardly in post-processing to allow for direct comparison between analyses.
\end{abstract}
   
\date{\today}

\maketitle
\section{Introduction}
\label{sec:intro}
Compact binary black holes (BBHs) emit gravitational waves (GWs) during the last stages of their coalescence.
During this process the system loses energy and angular momentum, causing the orbit to both shrink and progressively circularize \cite{Peters:1963ux}. This motivates the analysis of gravitational-wave signals with theoretical templates that are generated by waveform models using the quasi-circular approximation. However, recent studies highlight how accurate measurements of eccentricity can provide vital astrophysical information that could, for example, help discriminate between different binary formation channels \cite{Samsing:2017xmd,Rodriguez:2017pec,Fragione:2018yrb,Samsing:2020tda,Zevin:2021rtf,Tagawa:2020jnc}. Consequently, there has been a growing interest in analyzing the GW events detected by LIGO and Virgo with inspiral-merger-ringdown (IMR) waveform models that include eccentricity ~\cite{Gayathri:2020coq,Romero-Shaw:2022xko,Clarke:2022fma,Romero-Shaw:2021ual}. For example, the GW transient GW190521~\cite{LIGOScientific:2020iuh} has recently been analyzed under the hypothesis that it originated from a hyperbolic capture that resulted in a highly eccentric merger \cite{Gamba:2021gap}; other studies claim moderate eccentricity and spin-induced precession as evidence for dynamical formation \cite{Romero-Shaw:2020thy}, a possible head-on collision \cite{Bustillo:2020syj} or large eccentricity and strong spin-induced precession \cite{Gayathri:2020coq}.

One of the most promising approaches towards modelling the full GW signal emitted by compact binaries on arbitrarily eccentric orbits is the effective-one-body framework (EOB) \cite{Buonanno:1998gg,Buonanno:2000ef,Damour:2000we,Damour:2001tu}. Early attempts at incorporating eccentricity within the EOB framework were presented in \cite{Hinderer:2017jcs,Cao:2017ndf,Liu:2019jpg} but have seen numerous improvements over recent years \cite{Chiaramello:2020ehz,Nagar:2021gss,Placidi:2021rkh,Albertini:2021tbt,Albanesi:2022ywx,Albanesi:2022xge,Ramos-Buades:2021adz,Liu:2021pkr,Yun:2021jnh}. In addition to EOB, there have also been numerous developments using alternative approaches towards modelling the complete IMR signal from eccentric binaries, including Numerical Relativity (NR) surrogates~\cite{Islam:2021mha,Huerta:2017kez} and hybrid models that blend post-Newtonian (PN) evolutions with NR simulations~\cite{Ramos-Buades:2019uvh,Tiwari:2020hsu,Cho:2021oai,Chattaraj:2022tay}. A key limitation of these approaches, however, is that they are often constrained by the availability of accurate numerical relativity simulations that span the full parameter space and -- in the case of surrogates -- by the length of the simulations themselves, which often do not cover the early inspiral of the system.
Conversely, models based on analytical PN and scattering calculations \cite{Loutrel:2016cdw,Loutrel:2018ydu,Moore:2019xkm,Boetzel:2019nfw,Klein:2021jtd,Tucker:2021mvo}
can deliver representations of signals from long lasting inspirals, but they lack a description of the strong-field merger and are only valid for moderate eccentricities.

We are particularly interested in the \TEOBResumS{} model \cite{Nagar:2018zoe,Nagar:2020pcj,Riemenschneider:2021ppj} and the extension to eccentricity \cite{Chiaramello:2020ehz,Nagar:2021gss,Nagar:2021xnh} that is built on the idea of dressing the circular azimuthal component of radiation reaction with the leading-order (Newtonian) non-circular correction \cite{Chiaramello:2020ehz}. This approach has been subsequently extended to each multipole in the waveform and was further improved by incorporating higher order post-Newtonian information in an appropriately factorized and resummed form \cite{Placidi:2021rkh}. In particular, \cite{Placidi:2021rkh} extended the noncircular waveform up to 2PN using results that partially build on \cite{Khalil:2021txt}. Whilst several proposals exist for incorporating radiation reaction, a detailed survey of these schemes was conducted in \cite{Albanesi:2022ywx} concluding that the Newtonian factorization complete with 2PN corrections demonstrated the best agreement with results in the test-mass limit. This paradigm was further extended in \cite{Albanesi:2022xge}.  

In this work we focus on {\tt TEOBResumS} and study the performance of its circular and eccentric versions (\TEOBResumSGIOTTO{} and \TEOBResumSecce{}, respectively) when applied to GW parameter estimation. We do so with the aim of validating the model and gauging possible biases due to eccentricity (or lack thereof).
We dedicate special attention to the study of the quasi circular limit of \TEOBResumSecce{}, 
and investigate how its structural differences with respect to \TEOBResumSGIOTTO{} -- quantified in
terms of unfaithfulness against numerical relativity waveforms -- reflect on GW data analysis of synthetic signals and GW150914. We then introduce a method to estimate the eccentricity directly from GW observations and determine its evolution as a function of frequency. This procedure is efficient and suitable to be applied to any eccentric waveform model in post-processing. Furthermore it is advantageous for comparing different eccentric analysis of GW events.

The paper is organized as follows: In Sec.~\ref{sec:EOB} we summarize the main elements of the EOB
waveform model used here. In Sec.~\ref{sec:methods} we present a brief review of the elements of
Bayesian inference needed for our analysis. Section~\ref{sec:injections} is devoted to the validation of
the waveform model via specific injection and recovery analyses. The model is then used to analyze
GW150914 data in Sec.~\ref{sec:gw150914} and Sec.~\ref{subsec:ecc_calc} is dedicated to presenting our method to estimate the eccentricity evolution of a coalescing BBHs system in post-processing. Concluding remarks are reported in Sec.~\ref{sec:discussion}. Throughout we use $G=c=1$ unless stated otherwise.

\section{Quasi-circular and eccentric waveform model: \TEOBResumS{}}
\label{sec:EOB}

All analyses presented in this paper are performed with \TEOBResumS{}, either in its native quasi-circular version, \TEOBResumSGIOTTO{}~\cite{Riemenschneider:2021ppj}, or in its eccentric version,
\TEOBResumSecce{}~\cite{Nagar:2021gss}. 
In this section we describe in some detail the features of the two models, highlighting
their structural differences and quantifying their (dis-)agreement as measured by the unfaithfulness (or mismatch) defined as: 
\begin{equation}
\label{eq:unfaithfulness}
\bar{F} = 1-F= 1 - \underset{t_0, \phi_0}{\rm max}  \frac{\langle h_1 | h_2 \rangle}{\sqrt{\langle h_1 |h_1 \rangle \langle h_2 |h_2 \rangle}},
\end{equation}
where $(t_0, \phi_0)$ are the initial time and phase of coalescence, and $\langle h_1 |h_2 \rangle$ is the noise weighted inner product between two waveforms 
\begin{equation}
\label{eq:inner_prod}
\langle h_1 |h_2 \rangle = 4 \Re \displaystyle\int_{f_{\rm min}}^{f_{\rm max}} \frac{ \tilde{h}_1 (f) \tilde{h}^*_2 (f)}{S_n (f)} df,
\end{equation}
where $S_n (f)$ denotes the power spectral density (PSD) of the detector strain noise and $\tilde{h}_1 (f)$ and $\tilde{h}_2$ are the Fourier transforms of the time domain waveforms $h_1$ and $h_2$.

\subsection{Quasi-circular model: \TEOBResumSGIOTTO{}}
\label{subsec:giotto}

\TEOBResumSGIOTTO{} is a semi-analytical state-of-the-art EOB model
for spinning coalescing compact binaries~\cite{Damour:2014sva, Nagar:2015xqa, Nagar:2018zoe, Nagar:2019wds, Nagar:2020pcj, Riemenschneider:2021ppj}. The conservative sector of the model includes analytical Post-Newtonian (PN) information, resummed via Pad\'e approximants. 
Spin-orbit effects are included in the EOB Hamiltonian via two gyro-gravitomagnetic terms~\cite{Damour:2014sva},
while even-in-spin effects are accounted for through the centrifugal radius~\cite{Damour:2014sva}.
Numerical Relativity (NR) data is used to inform the model through an effective 5PN orbital parameter,
$a_6^c$, and a next-to-next-to-next-to leading order (NNNLO) spin-orbit parameter, $c_3$ \cite{Nagar:2018zoe}.
In the dissipative sector, waveform multipoles up to $\ell=8$ are factorized and resummed according to 
the prescription of~\cite{Nagar:2020pcj}. Next-to-quasicircular (NQC) corrections ensure a robust transition 
from plunge to merger, and a phenomenological NR-informed ringdown model completes the model for multipoles 
up to $\ell \leq 5$.
Although we focus here on BBH systems, we note that {\tt TEOBResumS-GIOTTO} can also generate waveforms
for binary neutron star coalescences, see~\cite{Nagar:2018zoe} and references therein.

Waveforms built from \TEOBResumSGIOTTO{} employing only the dominant multipole $\ell=|m|=2$ 
have been tested against the entire catalog of spin-aligned waveforms from the Simulating-eXtreme-Spacetimes 
(SXS) collaboration~\cite{SXS:catalog},
and were shown to be consistently more than $99\%$ faithful to NR~\cite{Riemenschneider:2021ppj}.
When higher modes are included in the dissipative sector of the model, the EOB/NR unfaithfulness always lies below the $0.3\%$ threshold when considering waveforms constructed only with the $\ell=|m|=2$ mode, and below $3\%$ for waveforms with modes up to $\ell=4$ if the system has total mass smaller than $120 M_{\odot}$~\cite{Nagar:2020pcj}.

\subsection{Eccentric model: \TEOBResumSecce{}}
\label{subsec:dali}

The eccentric generalization of $\tt TEOBResumS$, \TEOBResumSecce{}~\cite{Chiaramello:2020ehz, Nagar:2021gss}, 
builds on the features of the quasi-circular model detailed above but differs in few key aspects.
First, the quasi-circular Newtonian prefactor that enters the factorized waveform multipoles
is replaced by a general expression obtained by computing the time-derivatives of the Newtonian mass 
and current multipoles, as described in~\cite{Nagar:2021gss}. The same approach is implemented
for the azimuthal radiation reaction force.
Second, for eccentric binaries, the radial radiation reaction force $\mathcal{F}_r$ 
that contributes to the time evolution of the radial EOB momentum can no longer be 
neglected~\cite{Chiaramello:2020ehz}.
Third, the initial conditions must be specified in a different manner with respect to the quasi-circular case:
instead of employing the post-adiabatic procedure of \cite{Damour:2012ky}, \TEOBResumSecce{} computes adiabatic initial 
conditions and always starts the evolution of the system at the apastron, see Appendix~\ref{app:ics} for further details.
These conservative eccentric initial conditions, however, do not reduce to the quasi-circular initial conditions in the limit of 
small eccentricity. To partially correct for this issue, the quasi-circular initial conditions are manually imposed for $e_0 < 10^{-3}$. 
Finally, the values of $a_6$ and $c_3$ were modified in order to ensure that the model remains faithful to its quasi-circular limit \cite{Nagar:2021gss}.

\subsubsection{Quasi-circular limit of \TEOBResumSecce{}}
\label{subsubsec:qclimit}
All of the modifications above allow \TEOBResumSecce{} to provide waveforms and dynamics that are faithful to mildly eccentric 
SXS simulations \cite{Chiaramello:2020ehz, Nagar:2021gss}, scattering angle calculations \cite{Nagar:2021gss} and highly eccentric test-mass 
waveforms \cite{Albanesi:2021rby}. At the same time, however, because of these structural differences, the quasi-circular limit of the eccentric model \TEOBResumSecce{} does not exactly reduce to the \TEOBResumSGIOTTO{} model.
In order to quantify the agreement of \TEOBResumSecce{} with NR simulations and \TEOBResumSGIOTTO{}, respectively, we calculate the unfaithfulness defined in Eq.~\eqref{eq:unfaithfulness}. 

\begin{figure}[t]
\center
\includegraphics[width=\columnwidth]{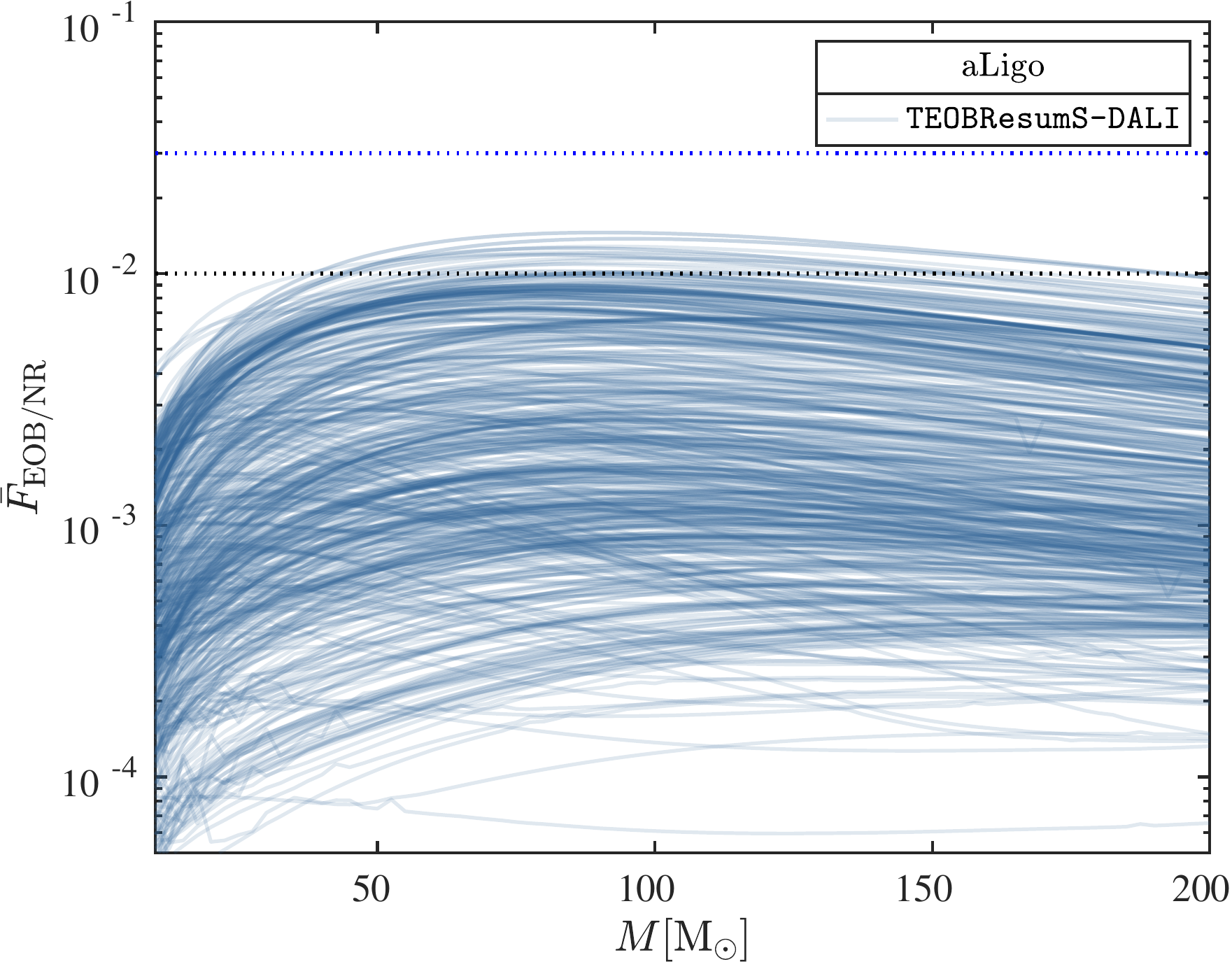}
\caption{\label{fig:barFeobnr}EOB/NR unfaithfulness using \TEOBResumSecce{} with $e_0^{\rm inj} = 10^{-8}$
over the non-precessing and non-eccentric SXS catalog. See text for more details.}  
\end{figure}
In Fig.~\ref{fig:barFeobnr} we show the unfaithfulness of \TEOBResumSecce{} against almost 
all\footnote{We exclude the following simulations due to large numerical errors: SXS:BBH:0002, SXS:BBH:1110, SXS:BBH:1141, SXS:BBH:1142}. non-eccentric, spin-aligned NR simulations in the SXS catalogue~\cite{Boyle:2019kee} using
the designed power spectral density (PSD) of Advanced LIGO~\cite{aLIGO:2020wna}. This figure complements, with many more simulations, 
Fig.~3 of~\cite{Nagar:2021gss}. Let us remind the reader that the corresponding plot for \TEOBResumSGIOTTO{} 
is centered around $10^{-3}$ with $\max(\bar{F}_{\rm EOBNR})\leq 9\times 10^{-3}$ with only a few outliers
above $3\times 10^{-3}$ (see Fig.~4 of~\cite{Riemenschneider:2021ppj}). We thus see here that \TEOBResumSecce{}
and \TEOBResumSGIOTTO{} are two EOB models, similarly informed by NR simulations, that perform differently 
with respect to quasi-circular NR simulations, though both are clearly below the usual threshold of $3\%$ unfaithfulness. 
It is therefore interesting to understand how this difference translates in terms of biases on parameters. 
This will be discussed in Sec.~\ref{sec:injections}. 

\section{Methods}
\label{sec:methods}

\subsection{Bayesian inference}
\label{subsec:bayes}
The measurement of the parameters that describe the GW emitting binary is carried out within the framework of Bayesian inference, which relies on Bayes' theorem \cite{Bayes:1764vd}
\begin{equation}
\label{eqn:bayes}
p(\Boldtheta|\Boldd,H)=\frac{p(d|\Boldtheta, H) \, p(\Boldtheta|H)}{p(\Boldd|H)},
\end{equation}
where $p(\Boldtheta|\Boldd,H)$ is the posterior probability of a set of parameters $\Boldtheta$ given the data $\Boldd$ assuming a specific model $H$, $p(\Boldtheta|H)$ is the prior, 
$p(\Boldd|\Boldtheta,H)$ is the likelihood 
and $p(\Boldd|H)$ is the evidence or marginalized likelihood. 
The evidence can be expressed as:
\begin{equation}
\label{eqn:evidence}
Z= p(\Boldd|H) = \int \, p(d|\Boldtheta,H) \, p(\Boldtheta|H) \Boldd\Boldtheta,
\end{equation} 
where the integral extends over the entire parameters space. The evidence assumes the role of an overall 
normalization constant but plays an important role in Bayesian model selection.
Given two competing hypotheses $H_A$ and $H_B$, the Bayes' factor is defined as the ratio of evidences
\begin{equation}
\mathcal{B}_A^B= \frac{p(\textbf{d}|H_B)}{p(\textbf{d}|H_A)} \, ,
\end{equation} 
where the hypothesis $H_B$ is favoured by the data over $H_A$ if $\mathcal{B}_A^B  > 1$. The expectation value of a parameter $\theta_i \in \Boldtheta$ can be estimated through the likelihood as
\begin{equation}
\label{eqn:mean}
E[\theta_i] = \int \theta_i \, p(\theta_i|\Boldd, H) d\theta_i,
\end{equation}  
where $p(\theta_i|\Boldd, H)$ is the marginalized posterior distribution for the parameter $\theta_i$. 

\subsection{Gravitational Wave Parameter Estimation}
\label{subsec:gwpe}
The GW signal emitted by an eccentric coalescing binary black hole system is fully described by $17$ parameters:
\begin{equation}
\label{eqn:thetacbc}
\Boldtheta_{\rm CBC} = \{m_1, m_2, \boldsymbol{\chi_1}, \boldsymbol{\chi_2}, D_L, \iota, \alpha, \delta, \psi, t_0, \phi_0, e_0, f_0\},
\end{equation}
where $m_{1,2}$ denotes the (detector-frame) masses of the two black holes such that $m_1 \geq m_2$, $\boldsymbol{\chi}_{1,2}$ are the dimensionless spin angular momenta vectors with three spatial components, $D_L$ is the luminosity distance to the source, $\iota$ is the inclination angle, $\{ \alpha, \delta\}$ are the right ascension and declination and define the sky location of the source, $\psi$ is the polarization angle, $\{ t_0, \phi_0 \}$ are the reference time and phase, and $\{e_0, f_0\}$ are the initial eccentricity magnitude and the average frequency between the apastron and periastron respectively.

In this work we utilize the \bajes{} package for Bayesian inference~\cite{Breschi:2021wzr} employing the nested sampling~\cite{Skilling:2006} algorithm \texttt{dynesty}~\cite{Speagle:2020} in order to extract the posterior probability density functions (PDFs) and to estimate the evidence.

\subsubsection{Likelihood}
\label{subsubsec:likelihood}

We are interested in the joint likelihood between $N$ detectors in a GW detector network 
\begin{equation}
p(\Boldd|\Boldtheta, H_S) = \prod_{i=1}^{N} p(\Boldd_{i}|\Boldtheta, H_S),
\end{equation} 
where $H_S$ denotes the hypothesis that the data contains a GW signal. Under the assumption of Gaussian, stationary noise that is uncorrelated between each detector, and assuming a time domain signal model $h \equiv h(t, \Boldtheta_{\rm CBC})$ and data set $ d \equiv d(t)$, the likelihood is given by
\begin{equation}
p(\Boldd|\Boldtheta_{\rm CBC}, H_S) \propto e^{-\frac{1}{2} \sum_{i=1}^N \langle h-d_i|h - d_i\rangle},
\end{equation}
where $\langle \cdot |\cdot \rangle$ is the noise-weighted inner product as defined in Eq.~\eqref{eq:inner_prod},
\begin{equation}
\langle h -d_i|h - d_i\rangle = 4 Re \int_0^{\infty} \frac{|\tilde{h} (f)-\tilde{d}_i(f)|^2}{S_n (f)} df,
\end{equation} 
where $S_n (f)$ is the PSD of the detector strain noise, and $\tilde{h} (f)$ and $\tilde{d}$ denote the Fourier transform of $h$ and $d$ respectively. 

\subsubsection{Priors}
\label{subsubsec:priors}

For the analyses presented in Sec.~\ref{sec:injections} we adopt priors that broadly follow \cite{LIGOScientific:2018mvr, Breschi:2021wzr} and are given as follows:
\begin{itemize}
\item The prior distribution for the masses is chosen to be flat in the components masses $\{ m_1, m_2\}$ and can be written in terms of the chirp mass $M_c = (m_1 m_2)^{3/5}/(m_1+m_2)^{1/5}$ and the mass ratio $q = {m_1}/{m_2} \geq 1$ as
\begin{equation}
\label{eqn:Mcprior}
p(M_c, q|H_S) = \frac{M_c}{\Pi_{M_c} {\Pi_{q}}} \left(\frac{1+q}{q^3}\right)^{2/5},
\end{equation}
where $\Pi_{M_c}$ and $\Pi_{q}$ are the prior volumes, as defined in Sec.V B of~\cite{Breschi:2021wzr} delimited by the prior bounds of $M_c$ and $q$. 
\item  
To aid the comparison with results from analyses that allow for precessing spins, we assume priors that correspond to the projection of a uniform and isotropic spin distribution along the $\hat{z}$-direction as proposed by Veitch~\cite{Lange:2018pyp,Breschi:2021wzr}: 
\begin{equation}
p(\chi_{i}|H_S) = \frac{1}{2 \chi_{\rm max}} \ln{\Bigg|\frac{\chi_{\rm max}}{\chi_{i}}\Bigg|},
\end{equation}
where $\chi_i$ is the magnitude of each black hole spin and $\chi_{\rm max}$ is the maximum spin magnitude.
\item The prior distribution for the luminosity distance $D_L$ is specified by a lower and an upper bound and its analytic form is defined by a uniform distribution over the sphere centred around the detectors: 
\begin{equation}
p(D_L|H_S) = \frac{3 D_L^2}{D^3_{\rm max}- D^3_{\rm min}}
\end{equation}
\item The prior distributions for $\alpha$ and $\delta$, defining the sky location, are taken to be isotropic over the sky with $\alpha \in [0, 2 \pi]$, $\delta \in [-\pi/2, +\pi/2]$ and 
\begin{equation}
p(\alpha, \delta|H_S) = \frac{\cos{\delta}}{4 \pi} .
\end{equation}
\item Analogously, for the inclination we have
\begin{equation}
p(\iota, H_S) = \frac{\sin{\iota}}{2},
\end{equation}
where $\iota \in [0, \pi]$.
\item For $\{\psi, t_0, \psi_0\}$, the prior distributions are taken to be uniform within the given bounds. 
\item The prior on $\{e_0, f_0\}$ are taken to be uniform or logarithmic-uniform within the provided bounds that are [0.001, 0.2] and [18, 20.5], respectively.
\end{itemize}

\section{Validation of the waveform model} 
\label{sec:injections}
In this section, we test the consistency of \TEOBResumSecce{} with \TEOBResumSGIOTTO{} (and vice-versa) by performing Bayesian 
inference on simulated GW signals (injections). The aim of this analysis is to give a more quantitative meaning to the standard
EOB/NR unfaithfulness figures of merit discussed above.
To do so, we inject mock signals into a zero noise realization with a signal-to-noise ratio (SNR) of 
$\sim 42$ in the Advanced LIGO and Advanced Virgo network. 
We employ the Advanced LIGO and Advanced Virgo design sensitivity PSDs~\cite{Abbott:2016xvh, aLIGO:2020wna, VIRGO:2014yos}.
All injections are performed at the same GPS time, $\rm t_{GPS} =1126259462.4$s.
We analyze segments of $8$s in duration with a sampling rate of $4096$Hz. We use \texttt{dynesty} to sample the 
posterior distributions, using the following setting: 3000 live points to initialise the MCMC chains, a maximum of $10^4$ MCMC steps, a stopping criterion on the evidence of $\Delta \ln Z = 0.1$, and we require five autocorrelation times before accepting a point. For all our analyses, we restrict the waveform model to only the $(2,|2|)$-mode, allowing us to analytically marginalize over the phase.

\subsection{Quasi-circular limit of the eccentric model} 
\label{subsec:qclimit}
\begin{figure}[t]
\center
\includegraphics[width=\columnwidth]{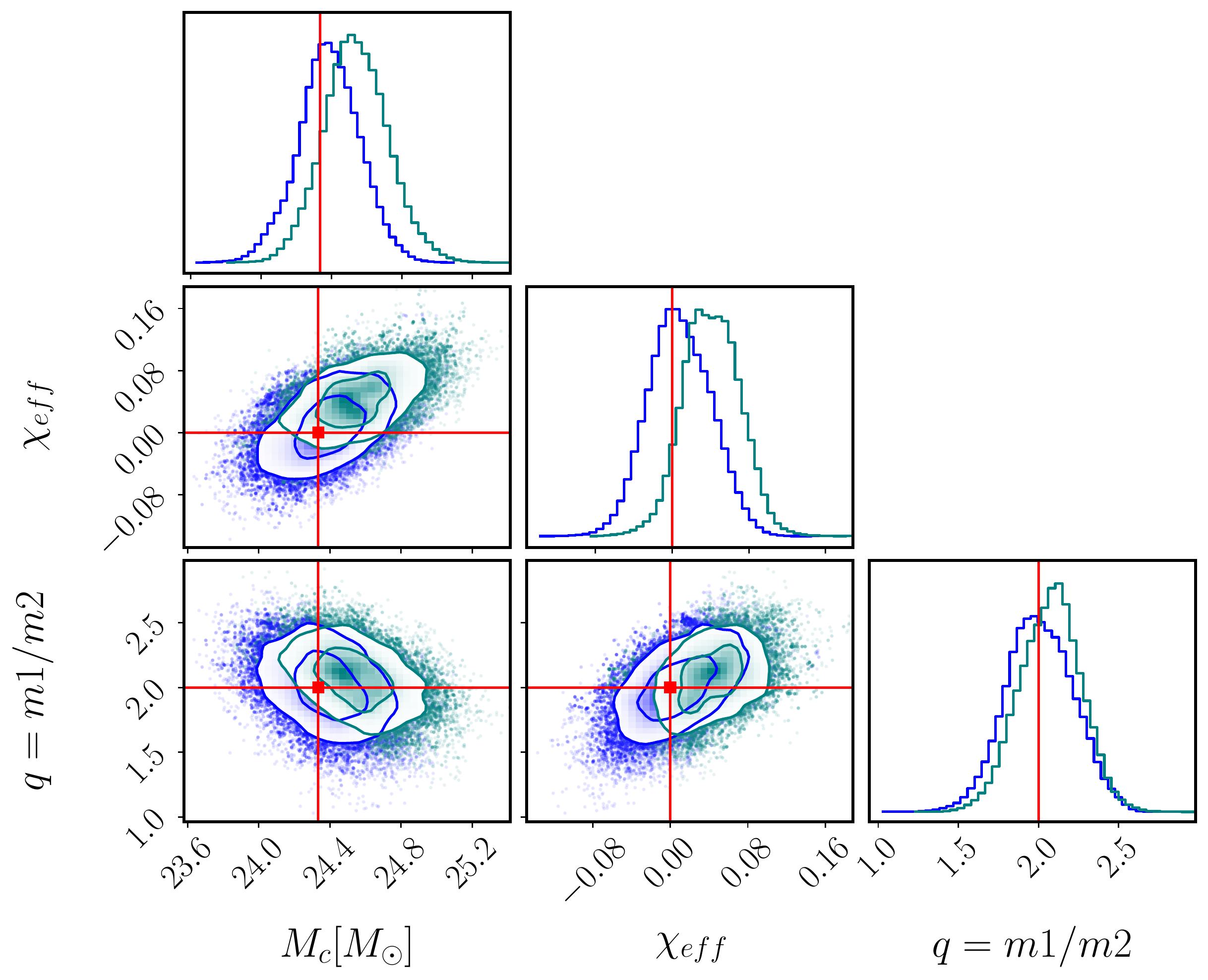}
\caption{\label{fig:corner_inj_circ}Testing the quasi-circular limit of \TEOBResumSecce{}. We inject a quasi-circular 
waveform generated with \TEOBResumSGIOTTO{} and recover it with either \TEOBResumSGIOTTO{} (blue) or 
with \TEOBResumSecce{} with fixed initial eccentricity at $e_0 = 10^{-8}$ (teal). The injected values are indicated by 
the solid lines. We find that the parameters recovered with \TEOBResumSecce{} are slightly biased. See text for discussion.}
\end{figure}

As mentioned above, \TEOBResumSecce{} is structurally different to the quasi-circular \TEOBResumSGIOTTO{} model. 
Moreover, despite having been informed by the same NR simulations, its unfaithfulness to NR is {\it larger} than 
that of \TEOBResumSGIOTTO{}. To better understand how this difference in the unfaithfulness translates into parameter 
biases, we perform an unequal mass injection in the quasi-circular limit, as detailed in Table~\ref{tab:circinj}.
More precisely, the injected waveform is generated with \TEOBResumSGIOTTO{} from a fixed initial frequency of 20 Hz, 
and it is recovered with either the same model (Prior 1) or with \TEOBResumSecce{} assuming a fixed initial 
eccentricity of $e_0 = 10^{-8}$ at 20 Hz (Prior 2). 
In Fig.~\ref{fig:corner_inj_circ} we show the one-dimensional and joint posterior distributions 
for $M_c$\footnote{We note that we quote the detector-frame chirp mass throughout the paper.}, $q$ and the effective spin
\begin{equation}
    \chi_{\rm eff} = \frac{m_1 \chi_{1z}+m_2 \chi_{2z}}{m_1+m_2},
\end{equation}
where the two spins are taken to be aligned along the $\hat{z}$-direction: $\chi_{1z}=\chi_{1}$ and $\chi_{2z}=\chi_{2}$.
The median values of $M_c$, $\chi_{\rm eff}$ and $q$ recovered with \TEOBResumSGIOTTO{} and \TEOBResumSecce{} are shown, with their 90$\%$ credibility interval, respectively in the first and second column of Table~\ref{tab:inj_circ_posteriors} in Appendix~\ref{app:tables}.
Comparing the results, we notice that the median values of the parameters recovered with \TEOBResumSGIOTTO{} are in good agreement with the injected ones, while those recovered with the quasi-circular limit of \TEOBResumSecce{} are slightly biased towards higher values. 
This is not surprising given the different analytical structures (dissipative sectors and NR-informed parameters) of the two models and the fact that \TEOBResumSecce{} is less NR-faithful than \TEOBResumSGIOTTO{} by, on average, one order of magnitude ($\sim 10^{-2}$ vs. $10^{-3}$) (see Fig.~\ref{fig:barFeobnr} and Fig.~4 of~\cite{Riemenschneider:2021ppj}). Moreover, when comparing the two models with each other, we also find an average unfaithfulness of $2-3\%$, which increases slightly with the total mass of the binary.

\begin{table}[t]
\ruledtabular
\begin{tabular}{c|c|c|c}
Parameter & Injected value & Prior 1 & Prior 2\\
\hline
$M_{c} (\rm M_{\odot})$ & 24.33 & $[18, 45]$ & $[18, 45]$  \\
$ q$ & 2 &  $[1, 3]$ & $[1, 3]$  \\
$\chi_{1}$ & 0 & $[-0.8, 0.8]$  & $[-0.8, 0.8]$\\
$\chi_{2}$ & 0 & $[-0.8, 0.8]$  & $[-0.8, 0.8]$\\
$ D_L \rm{(Mpc)}$ & 800 & $[50, 2000]$ & $[50, 2000]$ \\
$ \cos\iota$ & 0 & $[-1, 1]$  & $[-1, 1]$\\
$ \alpha \rm{(rad)} $ & 0.37 &  $[0, 2\pi]$ & $[0, 2\pi]$ \\
$\delta \rm{(rad)}$ & 0.81 &  $[-\pi/2, \pi/2]$ & $[-\pi/2, \pi/2]$\\
$\psi \rm{(rad)} $ & 0 & $\mathcal{U}(0, \pi)$ & $\mathcal{U} (0, \pi)$\\
$t_0 \rm{(s)}$ & 0 & $\mathcal{U}(-1, 1)$ & $\mathcal{U}(-1, 1)$\\
$\phi_0 \rm{(rad)}$ & 0 & -- & -- \\
$e_0$ & 0 & 0 & $10^{-8}$\\ 
$f_0 \rm{(Hz)}$ & 20 & 20 & 20\\
\hline
Model & \texttt{GIOTTO} & \texttt{GIOTTO} & \texttt{DALI} \\
\end{tabular}
\caption{\label{tab:circinj}
Parameters of the circular injection and two different priors. The prior distributions are described in Sec.~\ref{subsubsec:priors}. The sky location corresponds to the maximum sensitivity for the Advanced LIGO Hanford detector.}
\end{table}

\subsection{Testing the eccentric model}
\label{subsec:inj_ecc}
\begin{figure}[t]
\center
\includegraphics[width=\columnwidth]{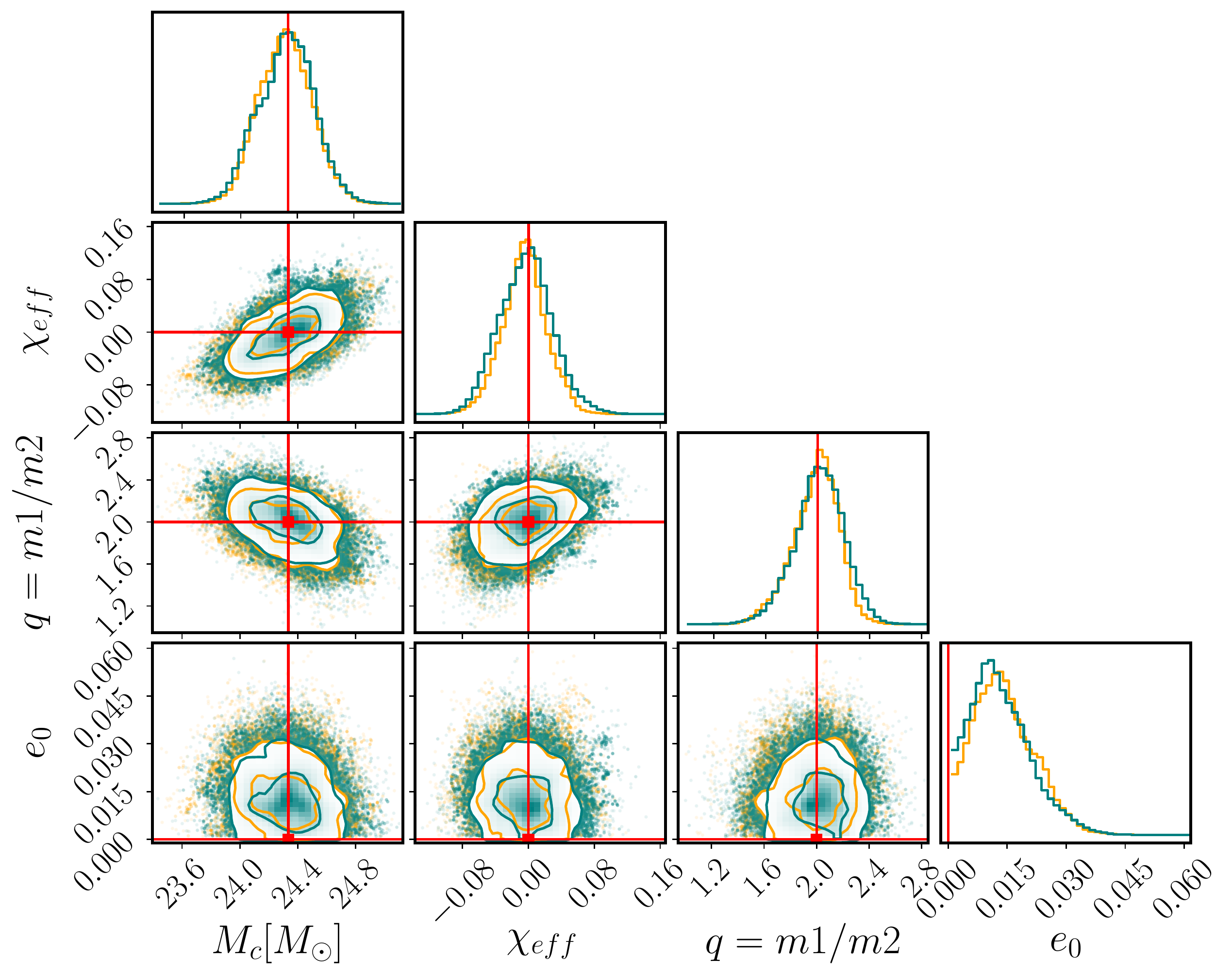}
\caption{\label{fig:corner_inj_circ_ecc_unif_bottom}
Injection with \TEOBResumSGIOTTO{} and recovery with \TEOBResumSecce{}.
The posterior distributions are obtained by sampling 
in $(e_0,f_0)$ (orange) or by sampling only in $e_0$ while keeping  $f_0$ fixed (teal). 
The injected values are represented by the solid lines. 
We do not find appreciable biases in the reconstructed parameters when sampling only in $e_0$.} 
\includegraphics[width=\columnwidth]{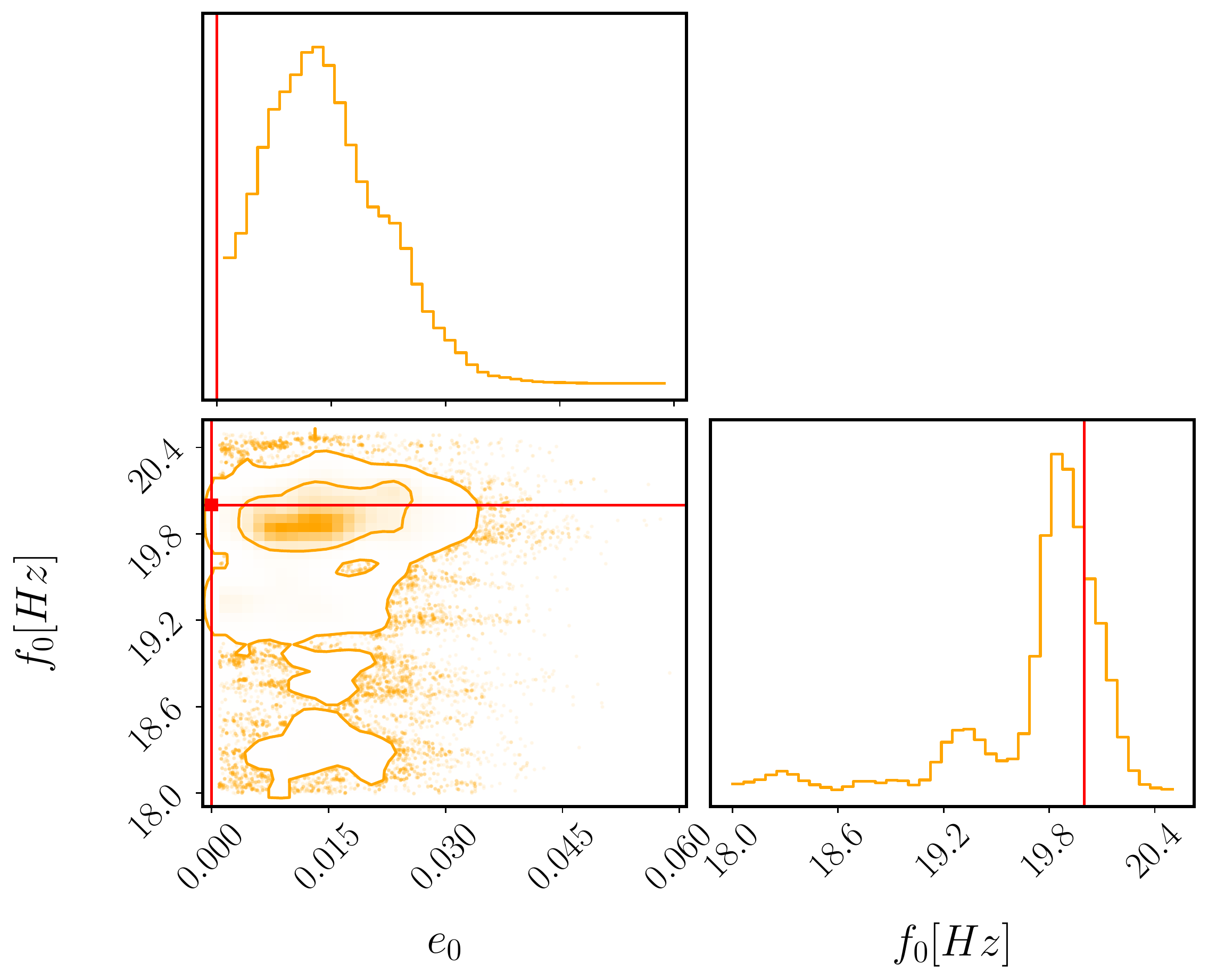}
\caption{\label{fig:corner_inj_circ_ecc_unif_up}Two-dimensional eccentricity and frequency posterior distributions for the same \TEOBResumSGIOTTO{} injection and recovery with \TEOBResumSecce{} as in Fig.~\ref{fig:corner_inj_circ_ecc_unif_bottom}. We do not observe any significant correlation between $e_0$ and $f_0$.}
\end{figure}

\begin{figure*}[t]
\center
\includegraphics[width=0.45\textwidth]{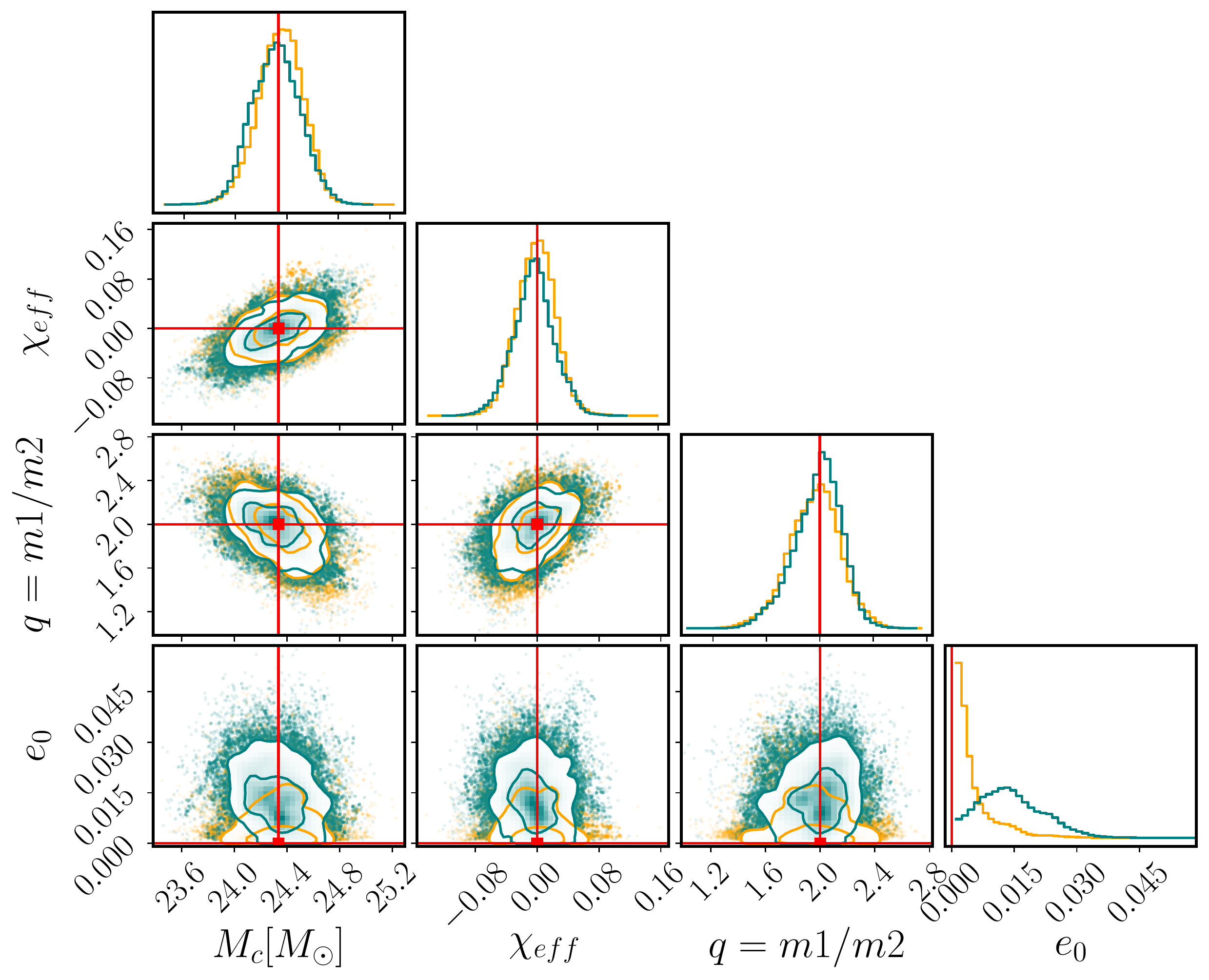}
\includegraphics[width=0.45\textwidth]{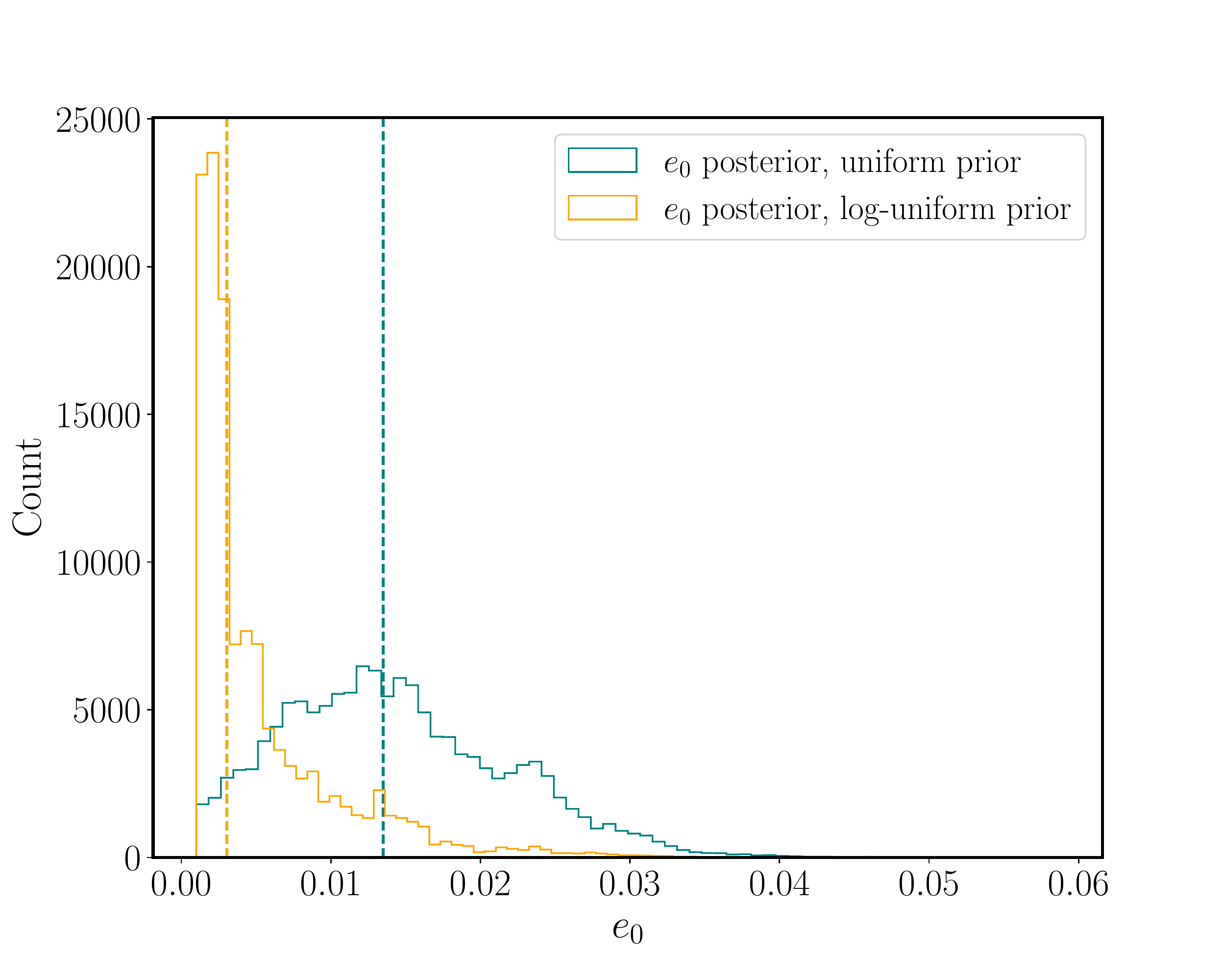}
\caption{\label{fig:corner_inj_circ_rec_ecc}Injection with \TEOBResumSGIOTTO{} and recovery with \TEOBResumSecce{}. 
Left: the posterior distribution for the $(M_c,q,\chi_{\rm eff})$. Right: the posterior distribution for the initial eccentricity using two different 
choices of priors: uniform (teal) and logarithmic-uniform (orange).}
\end{figure*}
\begin{figure*}[t]
\center
\includegraphics[width=0.45\textwidth]{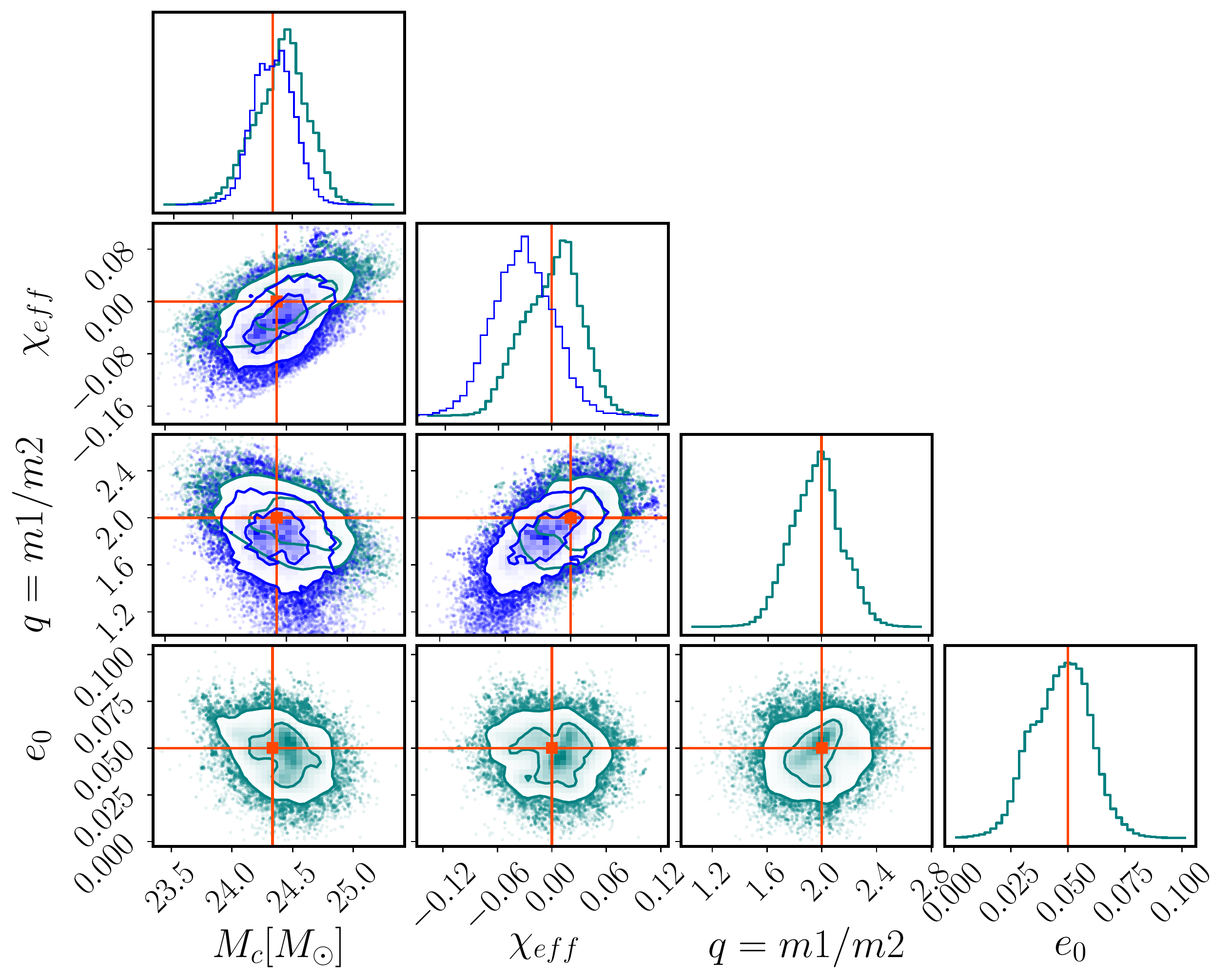}
\includegraphics[width=0.45\textwidth]{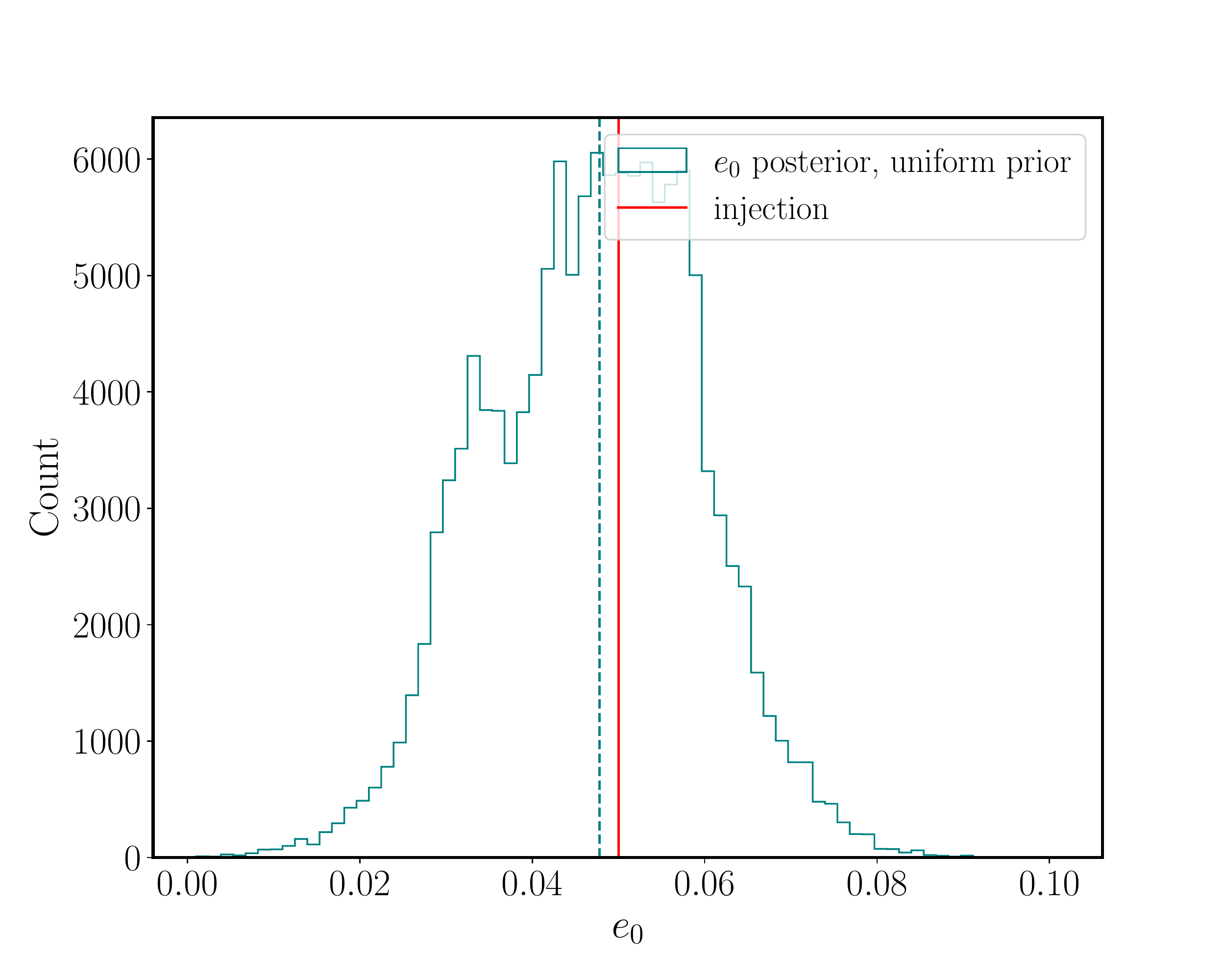}
\caption{\label{fig:corner_inj_ecc}
Injection with \TEOBResumSecce{} with fixed EOB eccentricity $e_0 = 0.05$ at $20$ Hz and recovery with \TEOBResumSGIOTTO{} (blue) or \TEOBResumSecce{} (teal).
When the analysis is performed with the latter, all recovered parameters look consistent with the injected ones.
Conversely, neglecting eccentricity leads to biases in the mass ratio and effective spin posterior distributions.}
\end{figure*}

In the EOB framework, the dynamics of a system of coalescing binaries is evolved from initial conditions. 
For the \TEOBResumSecce{} model, this is done by defining an initial eccentricity $e_0$ and an initial frequency 
$f_0$ and, through Eq.~\eqref{eq:freq0}-\eqref{eq:pr*0}, determining ($r_0$, $p_{\varphi}^0$, $p^0_{r*}$).
The degree to which the initial frequency $f_0$ has an impact on Bayesian inference and our ability to constrain this parameter from the observations is poorly understood. In previous similar analyses, comparable quantities, such as the argument of the periapsis or mean anomaly, have typically been ignored. However, recent studies \cite{Islam:2021mha,Romero-Shaw:2022xko} suggest that the mismatches can degrade as we vary these parameters for a given eccentricity. It is therefore useful to quantify the impact of $f_0$ on Bayesian inference. To do so we perform a non-eccentric injection with $e_0=0$ and $f_0=20$ Hz and recover with \TEOBResumSecce{} either sampling on $e_0$ and $f_0$ (Prior 1) or only on $e_0$ (Prior 2). The details of the injection and the priors are listed in Table~\ref{tab:eccfreq}. 
For the other parameters, the injected values and prior ranges are the same as in Table~\ref{tab:circinj}. 

 Figures~\ref{fig:corner_inj_circ_ecc_unif_bottom} and ~\ref{fig:corner_inj_circ_ecc_unif_up} show the one-dimensional and joint posterior distributions obtained with the two different priors. 
In Fig.~\ref{fig:corner_inj_circ_ecc_unif_bottom}, we show the posterior distributions for $M_c$, $\chi_{\rm eff}$, $q$ and $e_0$ obtained using the first prior choice (orange) and the second prior choice (teal). The median values, at $90 \%$ credibility, are shown in the third (Prior 1) and fourth (Prior 2) columns of Tab.~\ref{tab:inj_circ_posteriors} in App.~\ref{app:tables}.
We do not observe any significant differences between the two analyses and we find that the posterior on $f_0$ is weakly correlated with $e_0$ about its true value as can be seen from Fig.~\ref{fig:corner_inj_circ_ecc_unif_up}. This is in broad agreement with the conclusions of~\cite{Clarke:2022fma}, who found that the argument of periapsis is only likely to be resolvable for the loudest events. However, as also discussed in Refs. \cite{Clarke:2022fma,Romero-Shaw:2021ual, OShea:2021ugg}, we could potentially see biases if we fix $f_0$ to a frequency that effectively corresponds to the argument of the periapsis being out of phase with the true value. In the $e_0 \rightarrow 0$ limit, however, one may expect $f_0$ to become increasingly degenerate with the coalescence phase.

We note that although the injected value for $e_0$ is not contained within the priors, we do not see evidence that this impacts the inferred results. But we find a prior- dependence in the posterior of $e_0$ (see Fig.~\ref{fig:corner_inj_circ_rec_ecc} and the discussion below), in addition to the systematic differences between the two models in the circular limit already highlighted in Fig.~\ref{fig:corner_inj_circ}.

We next inject mock signals with two different values of $e_0$ and recover them using \TEOBResumSGIOTTO{} and \TEOBResumSecce{} respectively. The details of the injected values for $e_0$ and $f_0$ and their priors are described in Table~\ref{tab:ecc2}. The injected values and priors for the other parameters are the same as before as given in Table~\ref{tab:circinj}. 
Figure~\ref{fig:corner_inj_circ_rec_ecc} shows the one- and two-dimensional posterior distributions for $M_c$, $\chi_{\rm eff}$ and $q$ (left) and the one-dimensional posterior distribution for $e_0$ (right) for a non-eccentric injection recovered with \TEOBResumSecce{} with two different choices of prior distributions: logarithmic-uniform (teal), uniform (orange). 
The recovered median values corresponding to the Prior 1 (orange) are shown in the third column while the one corresponding to the Prior 2 (teal) are shown in the fifth column of Tab.~\ref{tab:inj_circ_posteriors} in App.~\ref{app:tables}. 
We observe that for eccentricities comparable to zero, the mass and spin measurements are robust and independent of the choice of eccentricity prior. In the right panel of Fig.~\ref{fig:corner_inj_circ_rec_ecc}, we observe that when using a logarithmic-uniform prior for the eccentricity, the recovered median value of the eccentricity is pushed to smaller values as a result of the priors. 

In Fig.~\ref{fig:corner_inj_ecc}, instead, we show the posterior distributions for an injection with $e_0 = 0.05$ (\TEOBResumSecce{}) and recovered with both the models, \TEOBResumSGIOTTO{} and \TEOBResumSecce{}. The median values of the parameters recovered with \TEOBResumSGIOTTO{} (orange) are indicated in the first column of Tab.~\ref{tab:inj_ecc_posteriors} in App.~\ref{app:tables}, while the ones recovered with \TEOBResumSecce{} (teal) are indicated in the second column of the same Table.
In the left figure, we observe a stronger correlation between mass and spin parameters when we recover with \TEOBResumSGIOTTO{}. Previous studies have pointed out correlations between the chirp mass, the effective inspiral spin and the eccentricity~\cite{Ramos-Buades:2019uvh, OShea:2021ugg, Romero-Shaw:2021ual}. As our recovery model neglects eccentricity, biases in the mass and spin parameters are anticipated to compensate for this. Lastly, we draw our attention on the right figure of the bottom panel, where it is shown how excellently the recovery of the eccentricity is accomplished pointing out the robustness and accuracy of the model. 

In terms of model selection, we find that for the non-eccentric injection, the recovery with \TEOBResumSGIOTTO{} is preferred with respect to the one with \TEOBResumSecce{} with an estimated logarithmic Bayes' factor of $\ln \mathcal{B}_{\rm circ}^{\rm ecc} \sim 9$. Similarly, for the eccentric injection, the eccentric model \TEOBResumSecce{} is preferred with respect to the quasi-circular model \TEOBResumSGIOTTO{} with an estimated logarithmic Bayes' factor of $\ln\mathcal{B}_{\rm ecc}^{\rm circ} \sim 5$ in the case of the uniform eccentricity prior, and $\ln\mathcal{B}_{\rm ecc}^{\rm circ} \sim 10$ when using the log-uniform prior. The difference in Bayes' factor between the two priors can be attributed to the $1/e_0$-scaling for the log-uniform prior, which \emph{a priori} favours smaller values of eccentricity.
The investigations presented in this section demonstrate that \TEOBResumSecce{} is a reliable waveform model to analyze spin-aligned, eccentric binaries.

\begin{table}[t]
\ruledtabular
\begin{tabular}{c|c|c|c}
Parameter & Injected value & Prior 1 & Prior 2 \\
\hline
$e_0$ & $0$ & $\mathcal{U}(0.001, 0.2)$ & $\mathcal{U}(0.001, 0.2)$ \\ 
$f_0 \rm{(Hz)}$ & 20 & $\mathcal{U}(18, 20.5)$ & 20 (fixed)\\
\hline
Model & --  & \texttt{DALI} & \texttt{DALI} 
\end{tabular}
\caption{\label{tab:eccfreq} Injected values for $e_0$ and $f_0$ and their priors. Two choices of recovery are made to perform this first testing analysis of \TEOBResumSecce{}. We choose to sample in both parameters in one case (Prior 1) and only in $e_0$ in the other case (Prior 2).}
\end{table}

\begin{table*}[!tp]
\begin{tabular}{c|c|c|c||c|c|c}
\hline
\hline
Parameter & Injection 1 & Prior 1 & Prior 2 & Injection 2 & Prior 1 & Prior 2 \\
\hline
$e_0$ & $0$ & $\mathcal{U}(0.001, 0.2)$ & Log-uniform$(0.001, 0.2)$ & 0.05 & $\mathcal{U}(0.001, 0.2)$ & 0 (fixed) \\ 
$f_0 (Hz)$ & 20& $\mathcal{U}(18, 20.5)$ & $\mathcal{U}(18, 20.5)$ & 20 & $\mathcal{U}(18, 20.5)$ & 20 (fixed)\\
\hline
Model & -- & \texttt{DALI} & \texttt{DALI} & \texttt{DALI} & \texttt{DALI}  & \texttt{GIOTTO} \\
\hline
\hline
\end{tabular}
\caption{\label{tab:ecc2} Second test of \TEOBResumSecce{} with an eccentric recovery. First column: injected values for $ e_0$ and $ f_0$ and their prior limits for an injection with $e_0=0$ injection recovered with \TEOBResumSecce{} with two different prior choices. Second column: injected values for $ e_0$ and $ f_0$ and their prior limits for an injection with $ e_0 = 0.05$ recovered with \TEOBResumSecce{} and \TEOBResumSGIOTTO{}.}
\end{table*}

\section{Analysis of GW150914}
\label{sec:gw150914}
In this section, we reanalyse GW150914 with the \TEOBResumSecce{} and \TEOBResumSGIOTTO{} waveform models. The strain data and PSDs are obtained from the GW Open Science Center~\cite{Trovato:2019liz}. We analyse an $8s$-long data stretch centered 
around the GPS time of the event $\rm t_{GPS} =1126259462.4 \, s$ sampled at a sampling rate of $\rm 4096 \, Hz$.
For the inference, we use \texttt{dynesty} choosing the same settings discussed in Sec.~\ref{sec:injections}.

\subsection{Quasi-circular analysis of GW150914}
\label{subsec:GW150914_circ}
\begin{figure}[h]
\center
\includegraphics[width=\columnwidth]{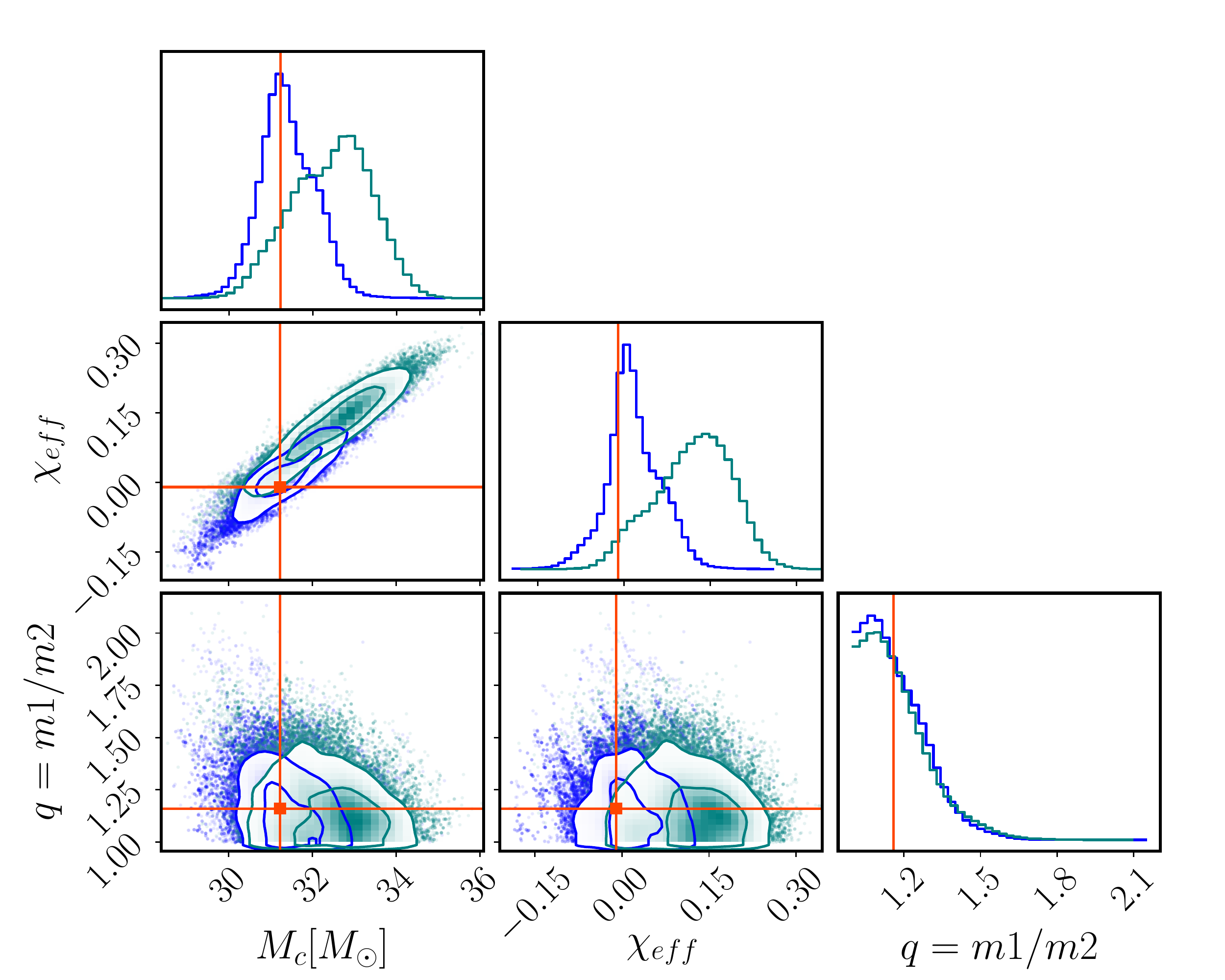}
\caption{\label{fig:corner_GW150914_qcirclim} One-dimensional and two-dimensional posterior distributions for $ M_c$, $ q$ and $\chi_{\rm eff}$ obtained with the quasi-circular model \TEOBResumSGIOTTO{} (blue) and the eccentric \TEOBResumSecce{} in the quasi-circular limit (i.e. $e_0$ fixed to $10^{-8}$ (teal)). The solid lines indicate the values from the quasi-circular analysis presented in GWTC-1~\cite{LIGOScientific:2018mvr}.
}
\end{figure}

First, we analyse GW150914 under the assumption of a quasi-circular binary black holes system. To do so, we perform two analyses,
either using \TEOBResumSGIOTTO{} or \TEOBResumSecce{}, fixing initial EOB eccentricity to $e_0 = 10^{-8}$, as described in Table~\ref{tab:circGW150914}. In both cases we recover a maximum likelihood SNR of 
$\sim 26$ corresponding to $\sim 20$ in LIGO-Hanford and $\sim 18$ in LIGO-Livingston.
In Fig.~\ref{fig:corner_GW150914_qcirclim} we show the marginalized one-dimensional and two-dimensional posterior distributions 
for $(M_c,\chi_{\rm eff},q)$ obtained with \TEOBResumSecce{} (teal) and \TEOBResumSGIOTTO{} (blue). The recovered median values are reported in the second and third column of Table~\ref{tab:GW150914_ecc_posteriors}. We observe that the values recovered with \TEOBResumSGIOTTO{} are consistent with the values for GW150914 reported in GWTC-1~\cite{LIGOScientific:2018mvr}, while the median values for the chirp mass and effective inspiral spin found with \TEOBResumSecce{} with fixed $e_0 = 10^{-8}$ are slightly higher in comparison to GWTC-1, 
but still consistent at the 90\% credible level. In terms of Bayes' factors we find that the analysis with \TEOBResumSGIOTTO{} 
is favored with a $\ln \mathcal{B}_{\rm ecc, 10^{-8}}^{\rm circ} \sim 1$. Based on the results for mock signals presented in Sec.~\ref{subsubsec:qclimit}, this is not surprising because of the structural difference between the two models 
and the influence of initial conditions on the quasi-circular limit as discussed extensively in Sec.~\ref{subsec:dali}.

\begin{table}[h]
\ruledtabular
\begin{tabular}{c|ccc}
\multicolumn{1}{c|}{Parameter} 
& \multicolumn{3}{c}{Prior} \\
\hline
$M_{c} (\rm M_{\odot})$ & $[12, 45]$ &&$[12, 45]$\\
$q$ & $ [1, 3]$ && $[1, 3]$ \\
$\chi_{1z}$ & $[-0.8, 0.8]$ && $[-0.8, 0.8]$  \\
$\chi_{2z}$ & $[-0.8, 0.8]$ && $[-0.8, 0.8]$  \\
$ D_L (\rm Mpc)$ & $[50, 2000]$ && $[50, 2000]$ \\ 
$\cos\iota$ & $[-1, 1]$  && $[-1, 1]$ \\
$\alpha (\rm rad) $ &$[0, 2\pi]$ && $[0, 2\pi]$  \\
$\delta (\rm rad) $ &  $[-\pi/2, \pi/2]$ && $[-\pi/2, \pi/2]$  \\
$\psi (\rm rad) $ & $\mathcal{U} (0, \pi)$ && $\mathcal{U}(0, \pi)$  \\ 
$t_0 (s)$ & $\mathcal{U}(-1, 1)$ && $\mathcal{U}(-1, 1)$  \\
$\phi_0 (\rm{rad})$ & -- && --\\
$ e_0$ & 0 (fixed) && $10^{-8}$ (fixed)\\
$ f_0 (\rm Hz)$ & 20 (fixed) && 20 (fixed) \\
\hline
Model & \TEOBResumSGIOTTO{} && \TEOBResumSecce{} \\
\end{tabular}
\caption{\label{tab:circGW150914} Choice of priors for the analysis of GW150914 to test the quasi-circular limit of \TEOBResumSecce{}. The prior distributions are described in detail in Sec.~\ref{subsubsec:priors}.}
\end{table}

\begin{table*}[!tp]
\begin{tabular}{c||c|c||c|c||c}
    \hline
    \hline
\multicolumn{1}{c||}{}   
&\multicolumn{5}{|c}{GW150914 Analysis}\\
\hline
Model &\TEOBResumSGIOTTO{}& \TEOBResumSecce{} &\TEOBResumSecce{}& \TEOBResumSecce{} &$\rm GWTC$-$1$\\
$e_0$-prior &$e_0 = 0$(fixed) & $ e_0 = 10^{-8}$ (fixed) & $\mathcal{U}$(0.001,0.2) & Log-uniform(0.001, 0.2) & -- \\
$f_0$-prior & $f_0 =$ 20 Hz (fixed) & $f_0=$ 20 Hz (fixed) & $\mathcal{U}$(18, 20.5) & $\mathcal{U}$(18, 20.5)& -- \\
\hline
$M_c (M_{\rm \odot})$ & $31.33^{+0.75}_{-0.52}$&$32.53^{+0.84}_{-1.08}$ & $31.54^{+0.92}_{-1.19}$ & $31.79^{+1.12}_{-0.94}$ & $31.23^{+1.08}_{-0.96}$  \\ 
$\chi_{\rm eff}$ & $0.01^{+0.05}_{-0.03}$&$0.13^{+0.06}_{-0.08}$ & $0.06^{+0.06}_{-0.10}$ & $0.08^{+0.08}_{-0.07}$ & $-0.01^{+0.12}_{-0.11}$  \\
$q$ & $1.14^{+0.14}_{-0.10}$ &$1.15^{+0.15}_{-0.10}$& $1.18^{+0.17}_{-0.11}$ & $1.21^{+0.19}_{-0.14}$  & $1.16^{+0.19}_{-0.11}$ \\
$e_0$ & -- & -- & $0.05^{+0.03}_{-0.02}$ & $0.02^{+0.03}_{-0.01}$ & -- \\
\hline
\end{tabular}
\caption{\label{tab:GW150914_ecc_posteriors} 
Results for the different analysis of GW150914 with \TEOBResumSGIOTTO{} or \TEOBResumSecce{}. The prior ranges for $e_0$ and $f_0$ for each analysis are indicated. We give the median values and symmetric 90\% credible interval for $M_c$, $\chi_{\rm eff}$ and $q$. Our results are contrasted by the values obtained from the non-eccentric, precessing analysis presented in GWTC-1~\cite{LIGOScientific:2018mvr} shown in the last column.}
\end{table*}

\subsection{Eccentric analysis of GW150914}
\label{subsec:GW150914_ecc}
\begin{figure}[!th]
\center
\includegraphics[width=\columnwidth]{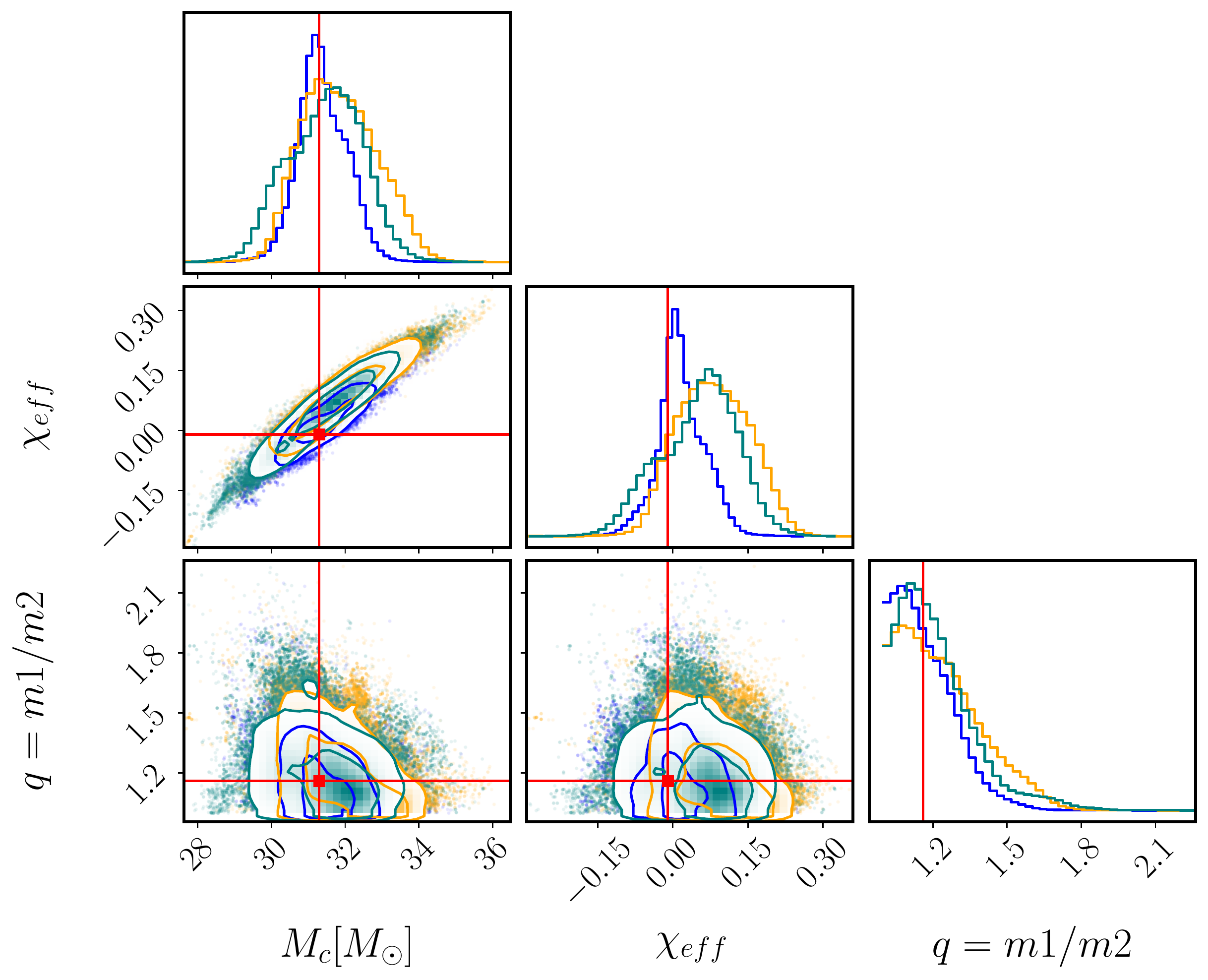}
\includegraphics[width=\columnwidth]{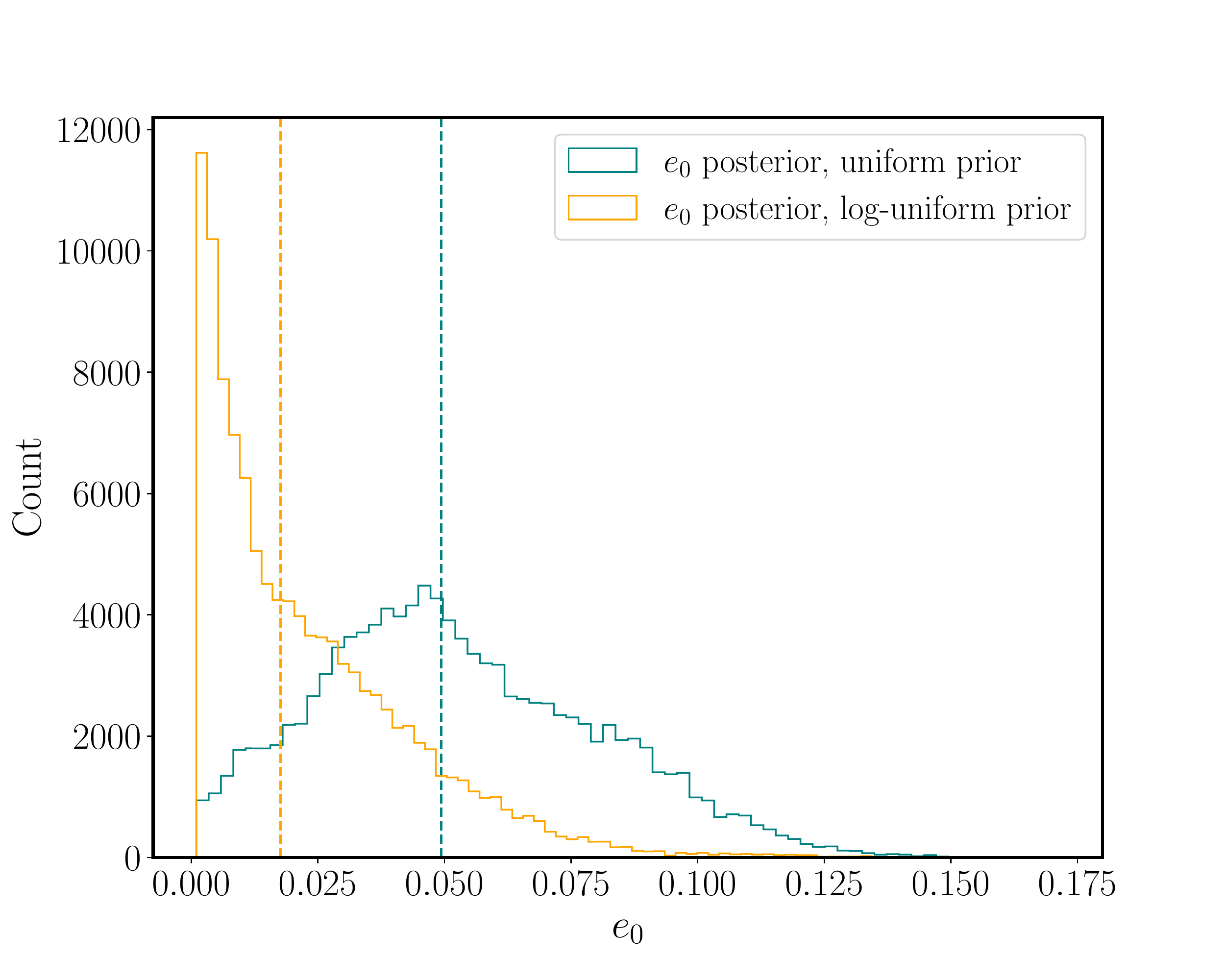}
\caption{\label{fig:GW150914_ecc_analysis}
Analyses of GW150914 with \TEOBResumSGIOTTO{} (blue) and \TEOBResumSecce{} with a uniform $e_0$-prior (teal) and a logarithmic-uniform $e_0$-prior (orange). Upper panel: Joint posterior distributions with 90$\%$ and 50$\%$ credibility interval and median values reported in GWTC-1 ~\cite{LIGOScientific:2018mvr} (solid lines). Bottom panel: Marginalised one-dimensional posterior distributions and median values of  $e_0$ (dashed lines) for the two eccentric analyses.} 
\end{figure}

Finally, we reanalyse GW150914 with the eccentric model \TEOBResumSecce{} sampling in both the initial eccentricity $e_0$ and $f_0$ (see Table~\ref{tab:GW150914_ecc_posteriors} for prior details). For the eccentricity we use two different priors: one uniform in $e_0$ and the other one logarithmic-uniform which occupies a larger prior volume at low eccentricities.
All other priors and settings are identical to the quasi-circular analysis of Sec.~\ref{subsec:GW150914_circ}.
Consistently with this, we estimate a network SNR of $\sim 26$ with $\sim 20$ in LIGO-Hanford and $\sim 18$ in LIGO-Livingston 
for the maximum likelihood parameters.
In Fig.~\ref{fig:GW150914_ecc_analysis} we show the one-dimensional and joint posterior distributions together with the median values reported in GWTC-1~\cite{LIGOScientific:2018mvr} 
or calculated from \cite{PE_samples} (solid lines).
The median values for $(M_c,\chi_{\rm eff},q)$ are given in Table~\ref{tab:GW150914_ecc_posteriors}.
The two eccentric analyses give consistent results for the mass and spin parameters, i.e. we do not 
find any appreciable difference between the results for the two different choices of the eccentricity prior. 
We do, however, find differences in the $e_0$ posterior under the two different prior assumptions as shown in the bottom panel of Fig.~\ref{fig:GW150914_ecc_analysis}. While both posteriors are consistent with small values of initial eccentricity, we find that the $e_0$-posterior peaks at $\sim 0.05$ for the uniform $e_0$-prior, which is in mild tension with other results~\cite{Romero-Shaw:2019itr, Iglesias:draft}. However, we note that this may be due to the uniform prior, which may not sufficiently explore low values of 
eccentricity. By contrast, when choosing the logarithmic-uniform prior, lower values of $e_0$ are preferred in full agreement with other analyses. We find that the maximum 90\% upper limit is $e_0 \lesssim 0.08$, which is consistent with the results based on NR simulations presented in~\cite{LIGOScientific:2016ebw}, where it was shown that the log-likelihood drops sharply as the eccentricity grows beyond $\sim 0.05$ at about $20$ Hz. 
For the other parameters (see Figs.~\ref{fig:corner_GW150914_intrinsic_ecc} and~\ref{fig:corner_GW150914_extrinsic_ecc} 
in Appendix~\ref{app:full_corner_GW150914}) we find broad agreement with the exception of the right ascension, where 
a different mode is preferred. In comparison to the quasi-circular analysis, the eccentric analyses give slightly 
higher median values for $M_c$ and $\chi_{\rm eff}$ in agreement with~\cite{Romero-Shaw:2019itr, Iglesias:draft}. 

In terms of model selection we find that \TEOBResumSGIOTTO{} is favoured over \TEOBResumSecce{} with an estimated Bayes' factor of $\ln\mathcal{B}^{\rm circ}_{\rm ecc} \sim 2$ irrespective of the prior. This is in agreement with the results reported 
in~\cite{ Romero-Shaw:2019itr}, but differs from the ones in \cite{Iglesias:draft}. However we note that 
Ref.~\cite{Iglesias:draft} uses higher order modes while in our analysis we only employ the dominant 
multipole $\ell = |m| = 2$ in the waveform. 
We conclude that, while the hypothesis of a quasi-circular BBH merger is preferred for GW150914, we cannot exclude a small value of eccentricity at $20$ Hz. All three analyses, however, give consistent results for the intrinsic parameters at 90$\%$ confidence. 
Our results are in agreement with previous analyses ~\cite{LIGOScientific:2018mvr, Romero-Shaw:2019itr, 
Romero-Shaw:2021ual}.

\section{Model-agnostic estimate of the Eccentricity Evolution}
\label{subsec:ecc_calc}

Bayesian inference allows us to determine the posterior distributions of binary parameters at a certain reference frequency. Certain parameters are, however, frequency dependent and hence change over time. One of these parameters is the eccentricity of the orbit, which decays due to the emission of GWs. In Sec.~\ref{subsec:GW150914_ecc} we determined the posterior distribution of the initial eccentricity $e_0$ of the EOB model measured at a (varying) reference average frequency $f_0$. We now devise a scheme to determine the evolution of the eccentricity as a function of frequency using a previously introduced eccentricity estimator~\cite{Mora:2002gf}. 
Gravitational radiation at future null infinity is expected to be manifestly gauge invariant, motivating the use of an estimator based on the relative oscillations in the gravitational-wave frequency. This mitigates against the contamination of eccentricity measurements through the use of gauge dependent quantities \cite{Mora:2003wt}.
This has the additional advantage of allowing for the direct comparison between different eccentric analyses, which often use different definitions of eccentricity~\cite{Knee:draft}\footnote{We remind the reader that in general relativity one does not have a unique, Newtonian-like definition of orbital eccentricity: due to periastron precession elliptic orbits do not generally close, even in the absence of dissipation caused by GW. Moreover, and most importantly, eccentricity is not a gauge invariant quantity, but rather it depends on the specific choice of coordinates. A detailed discussion on this topic can be found in e.g. \cite{Loutrel:2018ydu}.}.
Our scheme is computationally efficient and applicable to any eccentric waveform model in post-processing. A benefit of this way of estimating the eccentricity in post-processing is that it can be calculated directly from
the GW signal in contrast to definitions inferred from the dynamics \footnote{Nonethless, we note that since
we also have at hand the EOB dynamics, the same approach could be applied to the EOB orbital frequency.}. In addition, it also reduces to the Newtonian definition of eccentricity, even in the high eccentricity
limit ~\cite{Mora:2002gf,Ramos-Buades:2019uvh}.

To calculate the eccentricity evolution, we employ the eccentricity estimator first introduced by Mora et al.~\cite{Mora:2002gf}:
\begin{equation}
\label{eq:ecc_estimator}
e_{\omega}(t) = \frac{\omega_p(t)^{1/2} -\omega_a(t)^{1/2}}{\omega_p(t)^{1/2}+\omega_a(t)^{1/2}},
\end{equation}
where $ \omega_p(t)$ and $\omega_a(t) $ are fits to the GW frequency of the $(2,2)$-mode at the periastron and the apastron, respectively.
We note that this eccentricity estimator is also used in other works, e.g. either based on the orbital~\cite{Lewis:2016lgx,Ramos-Buades:2019uvh,Islam:2021mha} or the GW frequency~\cite{Chiaramello:2020ehz,Nagar:2021gss}.

To calculate $\omega_p(t)$ and $\omega_a(t)$, we first generate the \TEOBResumSecce{} waveform for each posterior sample 
and compute the GW frequency as $\omega(t) = \dot\phi(t)$, where $\phi(t)$ is the phase of the $(2,2)$-mode defined
as $h_{22}=A(t) e^{-i\phi(t)}$ with $A(t)$ being the amplitude of the waveform. We then identify the maxima (periastron) and the minima (apastron) of the {\it second time-derivative} 
of the GW frequency. We use the second derivative in order to amplify the peaks such that the identification of the 
maxima and minima is more robust for small eccentricities. 
%

\begin{figure}[t]
\center
\includegraphics[width=\columnwidth]{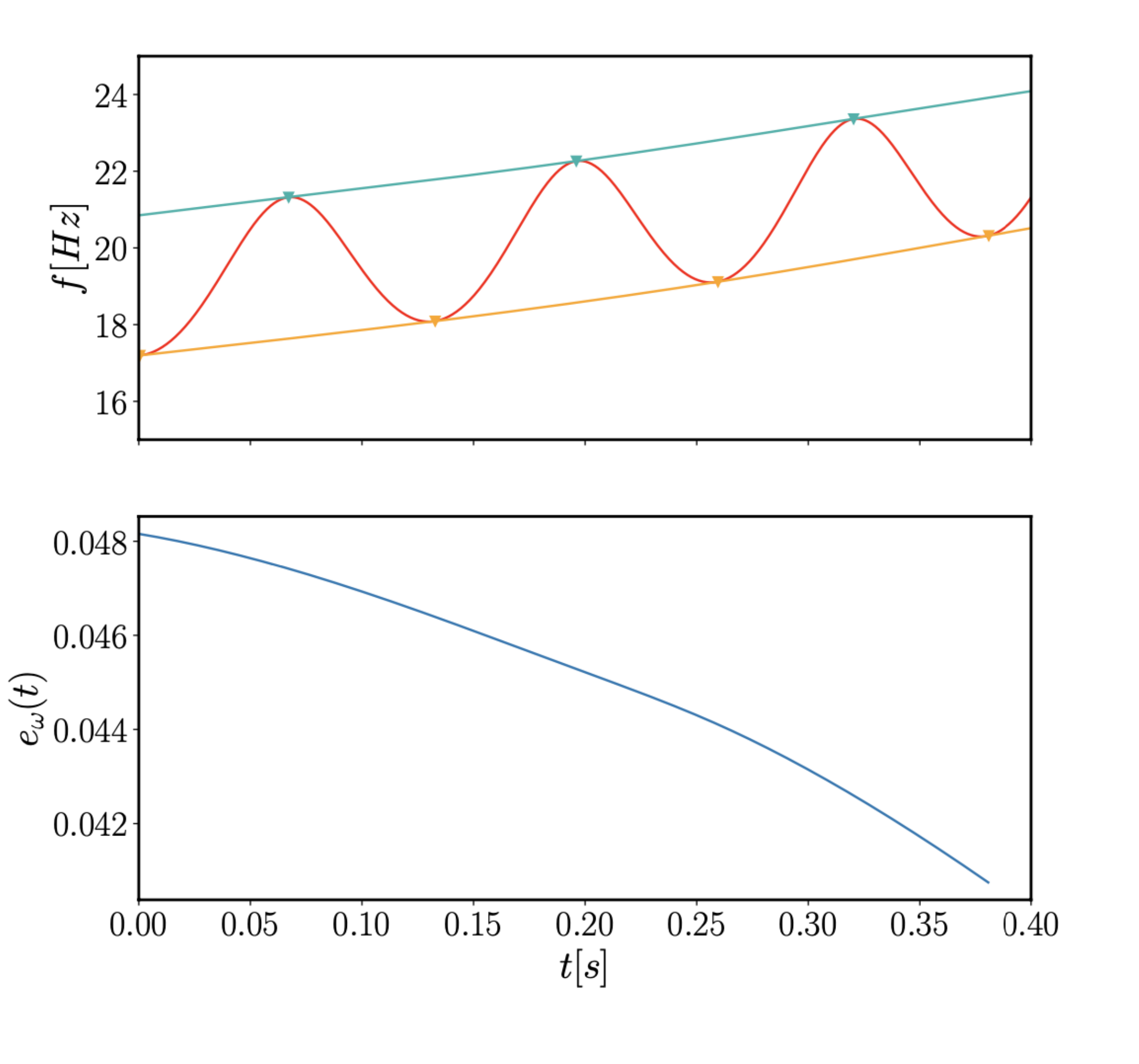}
\caption{\label{fig:goodfit}
Upper panel: Illustration of the fitting procedure to determine the maxima (teal) and minima (orange) of the GW frequency (red). Bottom panel: Evolution of the eccentricity $e_{\omega} (t)$ calculated using the 
method described in the text for a BBH with $M_c = 24.74$, $\chi_{\rm eff} = 0$ and $q = 1.5$. }
\end{figure}
Once the minima and maxima are identified, we fit $f(t) = \omega (t)/(2 \pi)$ using cubic spline interpolation.
An example of this is shown in the upper panel of Fig.~\ref{fig:goodfit}, where the red curve shows the GW frequency with clearly visible eccentricity-induced oscillations and the green and orange curves show the fits to the maxima and minima respectively. From Eq.~\eqref{eq:ecc_estimator} we calculate $e_{\omega} (t)$ for each posterior sample to find the corresponding eccentricity evolution, as shown in the bottom panel of Fig.~\ref{fig:goodfit}. We note that the eccentricity estimated at the initial time $e_{\omega}(t=0)$ can differ from the initial EOB eccentricity $e_0$ defined by the EOB dynamics as the eccentricity at the average frequency between apastron and periastron, as explained by Eq.~\eqref{eq:r0}. 

Since we are interested in determining how the eccentricity decays as the GW frequency increases towards merger, 
we need to map $t\rightarrow f$. Due to the non-monotonic behavior of the GW frequency, such a mapping is not 
unique and hence we introduce the average GW frequency $\bar{f}(t)$ instead:
\begin{align}
\bar{f}(t) = \frac{1}{2} \left(f_p (t) + f_a (t) \right),
\end{align}
where $f_p (t) = \omega_p (t) / (2 \pi)$ and $f_a (t) = \omega_a (t) / (2 \pi)$, and use linear interpolation 
to infer the eccentricity as a function of $\bar{f}$ throughout the inspiral.

As we mentioned before, this method benefits of the fact that it allows the eccentricity to be calculated directly from the GW signal and it  reduces to the Newtonian definition of eccentricity, even in the high eccentricity 
limit, however, the method also has some limitations.
A caveat to the correct calculation of $e_\omega(t)$ is, in fact, that it requires the inspiral to be sufficiently 
long such that many periastron and apastron peaks can be resolved. In particular, for short waveforms where we only 
have one or two maxima and minima available, this method is expected to become inefficient and inaccurate~\cite{Ramos-Buades:2019uvh}. 
A way to circumvent this situation is to generate the EOB waveforms from a lower starting frequency but at the cost of 
increasing the waveform generation time and hence the time taken for a Bayesian inference run to complete. 
Similarly, in the low-eccentricity limit, we may also expect peak-finding algorithms to become numerically unstable. 
While strategies to amplify the peaks, such as the use of the second derivative of the frequency, help to isolate 
the stationary points, in practice we found that the peaks can still be poorly resolved for a small subset of the samples. 
However, by cutting the frequencies at sufficiently small times ($t=0.4$ s), we found the eccentricity estimator to be numerically robust with only a small percentage of samples $(\lesssim 0.03   \%)$ potentially suffering from pathologies. For those samples, we can adjust the cutoff time/frequency to produce an estimate of the eccentricity.    

\begin{figure}[t]
\center
\includegraphics[width=\columnwidth]{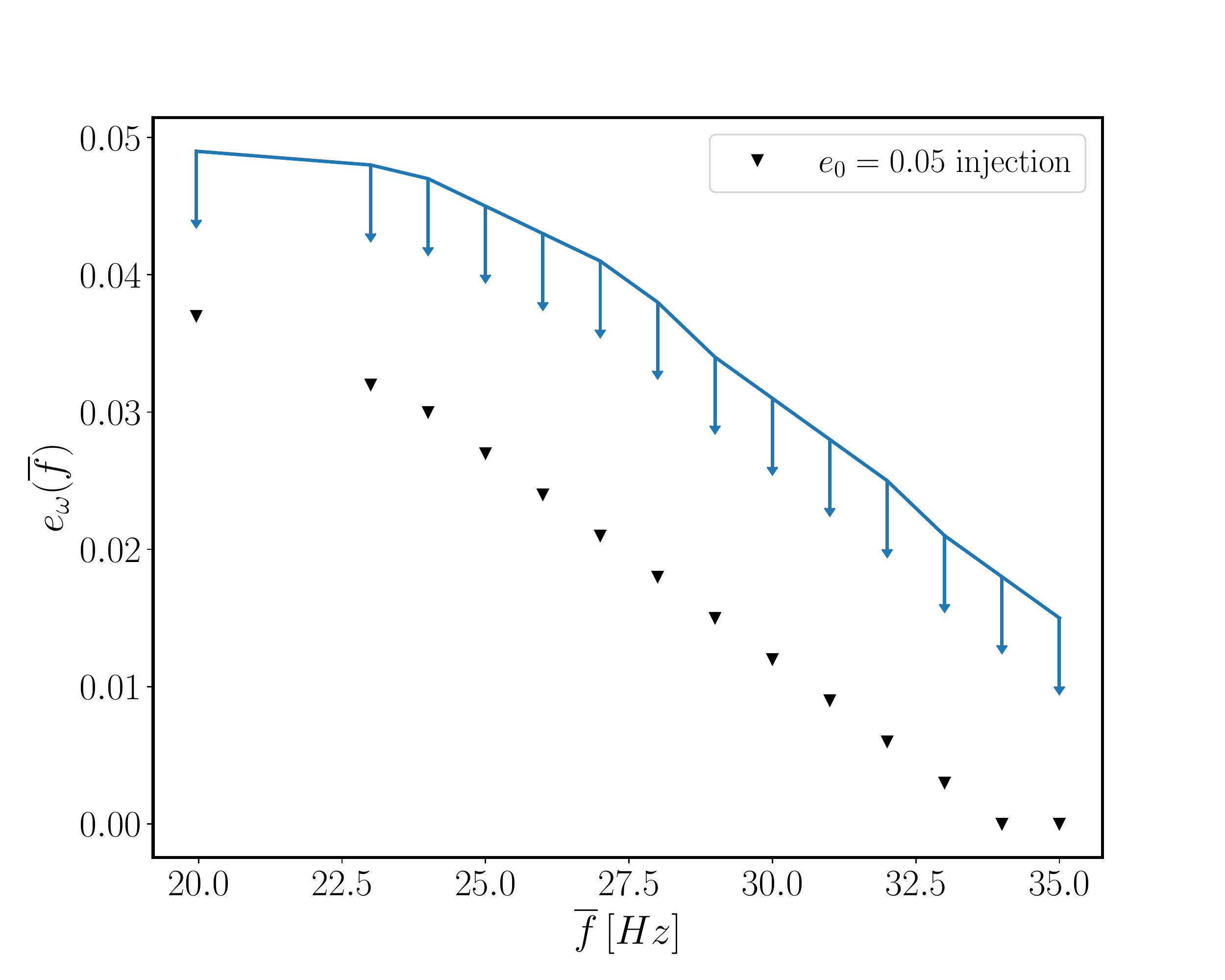}
\caption{\label{fig:e_perc_inj} 
Upper limit of the 90$\%$ credibility interval for the estimated eccentricity evolution $e_{\omega}(\bar{f})$ for an injection with $e_0 = 0.05$. The upper limit is calculated estimating $e_{\omega}(\bar{f})$ for all the posterior samples, interpolating it at different values of $\bar{f}$ and then taking the 90 $\%$ credibility interval of the of the data. The black triangles represent the injection. We note that the estimated initial eccentricity is slightly lower than $e_0 = 0.05$, where $e_0$ is defined from the EOB dynamics. } 
\end{figure}

In Fig.~\ref{fig:e_perc_inj} we show the 90\% upper limit of the eccentricity evolution $e_{\omega}(\bar{f})$ as a function of the average frequency for the simulated eccentric signal with $e_0 = 0.05$ and $ f_0 = 20$Hz, as discussed in Sec.~\ref{subsec:inj_ecc}. In addition, we also show the eccentricity evolution for the injected waveform itself (black triangles). We see that it is always contained within the $90\%$ upper limit. 

\begin{figure}[t]
\center
\includegraphics[width=\columnwidth]{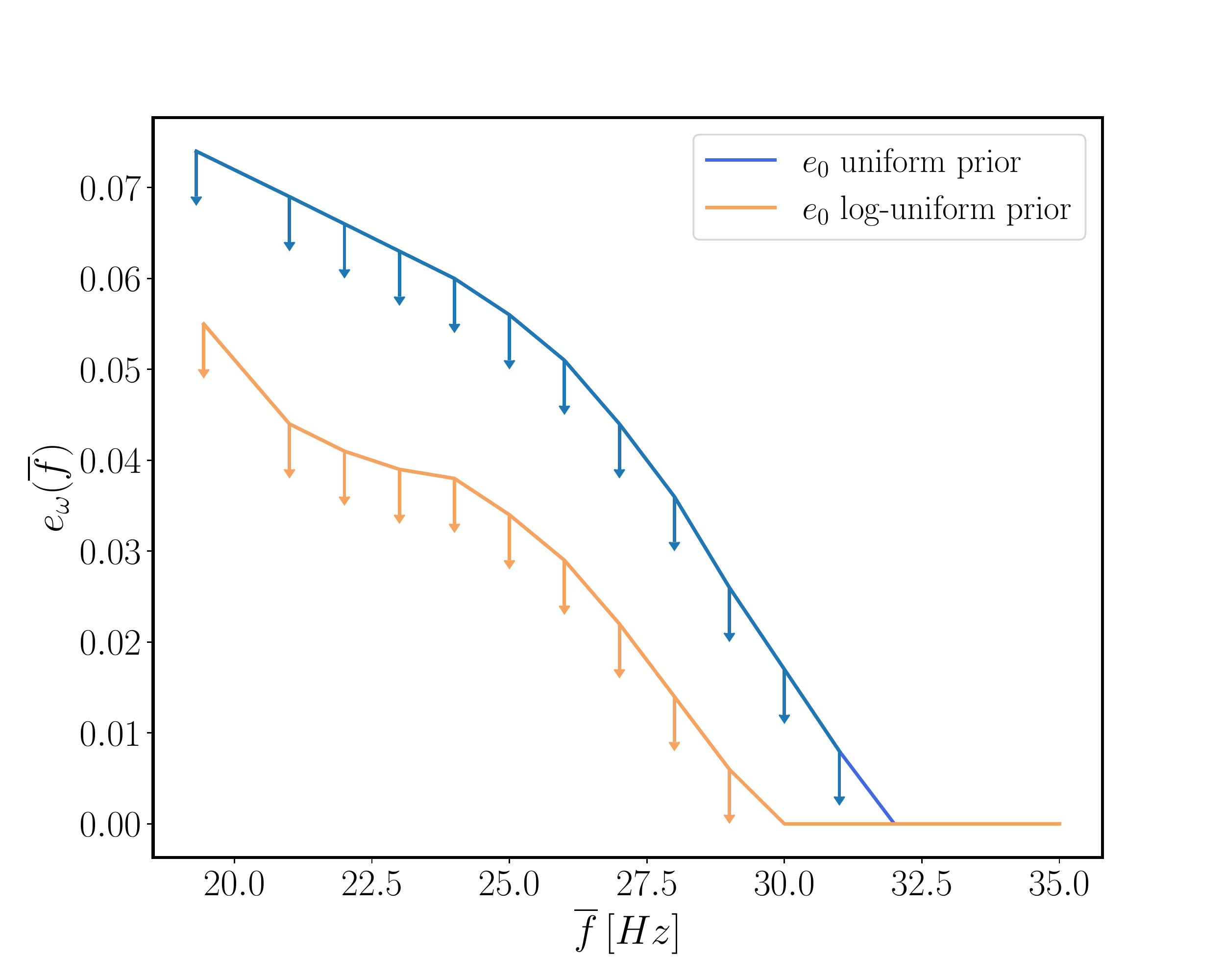}
\caption{\label{fig:e_f_unif} Upper limit of the 90$\%$ credibility interval for the estimated eccentricity evolution $e_{\omega}(\bar{f})$ for the two eccentric analyses of GW150914 with \TEOBResumSecce{}. The upper limit is calculated estimating $e_{\omega}(\bar{f})$ for all the posterior samples, interpolating it at different values of $\bar{f}$ and then taking the 90 $\%$ credible interval of the data. This result is agreement with previous results~\cite{LIGOScientific:2016ebw}.}
\end{figure}

Finally, we apply the same method to calculate the eccentricity evolution for GW150914 from the posterior samples obtained using the eccentric \TEOBResumSecce{} model as outlined in Sec.~\ref{subsec:GW150914_ecc}. Figure~\ref{fig:e_f_unif} shows the $90 \%$ upper limit of $e_{\omega}(\bar{f})$ obtained for the uniform $e_0$-prior distribution (blue) as well as for the log-uniform $e_0$-prior distribution (orange). We obtain an upper limit of $e_{\omega}(\bar{f})$ at $\sim$ 20 Hz of $\sim 0.075$ for the analysis with the uniform $e_0$-prior and $\sim 0.055$ for the analysis with the logarithmic-uniform $e_0$-prior. This is comparable with Fig.~7 of~\cite{LIGOScientific:2016ebw} where it was found that GW150914 is unlikely to have an eccentricity higher than $\sim$ 0.05 at about 20 Hz at $90 \%$ credibility.
We also see that while we cannot exclude small values of eccentricities at low frequencies, once an average frequency of $\sim 30$ Hz is reached, any residual eccentricity $e_\omega (\bar{f})$ can no longer be distinguished from zero.

\section{Discussion}
\label{sec:discussion}
In this work we present a Bayesian validation of the \TEOBResumSecce{} waveform model~\cite{Nagar:2021gss} 
for eccentric coalescing binary black holes with aligned spins, a fully Bayesian reanalysis 
of GW150914 and a systematic method to estimate the eccentricity in post-processing. Our study explores 
the potential of \TEOBResumSecce{} and allows us to test its reliability. 
Our work is an extension of our previous study~\cite{Nagar:2021gss} and demonstrates the efficacy 
of the model in distinguishing between circular and eccentric GW signals.
In particular, we find that the differences between the quasi-circular limit of \TEOBResumSecce{}
and its quasi-circular companion \TEOBResumSGIOTTO{} are relevant, and lead to clear (though small) 
biases in the recovered parameters. We attribute these biases to differences between the two models 
in both the dynamics (and especially in the radiative sector) and the waveform itself. 
When performing parameter estimation with small fixed eccentricity\footnote{We note that if the initial
eccentricity is sufficiently small the setup of the initial data is identical in both models.} 
this results in appreciable differences in the posteriors of numerous parameters. 
This indicates that the original \TEOBResumSecce{} model needs improvements, notably to recover
a quasi-circular limit that is as accurate as the one of \TEOBResumSGIOTTO{}. Some work in this
direction has been done~\cite{Nagar:2021xnh} (see in particular Fig.~8 therein) but more investigations
are needed to improve the model in the nearly equal-mass regime\footnote{We also note that the 
{\tt TEOBResumS} strategy is rather different from the one followed by the {\tt SEOBNRv4EHM } 
model~\cite{Ramos-Buades:2021adz} that substantially limits itself at changing initial conditions, 
without touching the structural elements of the dynamics. Although this choice guarantees, by construction,
an excellent quasi-circular limit, it introduces inaccuracies for eccentric dynamics, as highlighted 
in Ref.~\cite{Albanesi:2022ywx}}.

After testing \TEOBResumSecce{} for quasi-circular binaries, we validate the model on injections with nonzero initial
eccentricity. In particular we find that \TEOBResumSecce{} excellently recovers the injected value of 
eccentricity. In addition, we quantify the impact of eccentricity on the estimation of the intrinsic parameters 
of the binary: notably, we observe that the correlations between parameters became less strong when 
introducing eccentricity. If neglecting eccentricity, however, we see biases in the mass and spin parameters to compensate for it.

We then perform Bayesian inference with \TEOBResumSecce{} on the first GW event, GW150914. We find that the circular analysis is preferred with respect to the eccentric ones with $\ln\mathcal{B}^{\rm circ}_{\rm ecc}\sim 2$. However, we also find that we cannot exclude small values of eccentricities at low frequencies, and that once an average frequency of $\sim 30$ Hz is reached, any residual eccentricity becomes indistinguishable from zero. 

Lastly we perform the calculation of the eccentricity evolution using an eccentricity estimator
deduced from the instantaneous GW frequency. After testing the calculation on mock signals, we 
apply the method to the data of GW150914 finding that, at about 20 Hz, the maximum eccentricity allows for 
the system is $\sim 0.075$ for a uniform prior and $\sim 0.055$ for a logarithmic-uniform prior on the initial eccentricity. 
This is quantitatively comparable with the findings of~\cite{LIGOScientific:2016ebw}.
In the late stages of the preparation of this
manuscript we became aware of related but independent work
on eccentricity definitions~\cite{Shaikh:inprep}.

Given current BBH merger rate estimates~\cite{LIGOScientific:2021psn} and the sensitivity of the LIGO-Virgo-KAGRA detector network~\cite{LVK:observingscenario}, future detections of eccentric binaries will significantly constrain the lower limit of mergers that result from clusters and other dynamical channels~\cite{Zevin:2021rtf}. The possibility of several eccentric BBH candidates~\cite{Romero-Shaw:2022xko, Iglesias:draft} makes it crucial to have a reliable method to infer the eccentricity directly from observations. For the first time we present a systematic method to infer the eccentricity evolution directly from observations of GWs from coalescing BBHs that can be used in the future to robustly measure the eccentricity and make meaningful comparisons between different models. 

\begin{acknowledgments}
We thank the LIGO-Virgo-KAGRA Waveforms Group and, in particular, Vijay Varma, Antoni Ramos-Buades, Md Arif Shaikh, and Harald Pfeiffer for useful discussions and comments on the manuscript.
We also thank Alan Knee for helpful discussions during the development of this work.
A.~B. is supported by STFC, the School of Physics and Astronomy at the University of Birmingham and the Birmingham Institute for Gravitational Wave Astronomy. A.~B. acknowledges support from the Erasmus Plus programme and Short-Term Scientific Missions (STSM) of COST Action PHAROS (CA16214) for the first part of the project when she was visiting the Theoretisch-Physikalisches Institut in Jena.
R.~G. and M.~B acknowledge support from the Deutsche Forschungsgemeinschaft (DFG) under Grant No. 406116891 within the Research Training Group RTG 2522/1. 
P.~S. and G.~P. acknowledge support from STFC grant No. ST/V005677/1. Part of this research was performed while G.~P. and P.~S. were visiting the Institute for Pure and Applied Mathematics (IPAM), which is supported by the National Science Foundation (Grant No. DMS-1925919). G.P. is grateful for support from a Royal Society University Research Fellowship URF{\textbackslash}R1{\textbackslash}221500. 
S.~B. and M.~B. acknowledge support by the EU H2020 under ERC Starting Grant, no.~BinGraSp-714626. P.~R. aknowledges support by the Fondazione Della Riccia.
Computations were performed on the Bondi HPC cluster at the Birmingham Institute for Gravitational Wave Astronomy 
and the ARA supercomputer at Jena, supported in part by DFG grants
INST 275/334-1 FUGG and INST 275/363-1 FUGG and by EU H2020 ERC Starting Grant, no.~BinGraSp-714626. 
The waveform model used in this work is \TEOBResumS{} and is publicly developed and available at \url{https://bitbucket.org/eob_ihes/teobresums/}.
\noindent
Throughout this work we employed the commit 0f19532 of the eccentric branch. 
To perform Bayesian inference we used the \bajes{} software publicly available at \url{https://github.com/matteobreschi/bajes}. In this work we used the version available at \url{https://github.com/RoxGamba/bajes/commits/dev/teob_eccentric} employing the commit b3ad882. 
\noindent
This manuscript has the LIGO document number P2200219. 
\end{acknowledgments}

\appendix
\section{Quasi-circular and eccentric initial conditions}
\label{app:ics}
For quasi-circular binaries, {\tt TEOBResumS} applies Kepler's law to the 
initial frequency of the orbit to compute the initial separation $r$. Then, 
the initial values of the EOB angular and radial momenta $p_{\varphi}, p_{r_*}$ are estimated
via an iterative process (known as post-adiabatic expansion, ``PA" henceforth) in which 
the right-hand side of the Hamilton equations is solved analytically under the assumption
that $p_{r_*} \sim 0$~\cite{Damour:2012ky,Nagar:2018gnk}. At zeroth PA order, one assumes that $p_{r_*} = 0$ exactly. Then,
by evaluating $\partial_r \hat{H}_{\rm EOB} = 0$ one can analytically find the circular 
angular momentum $j_0(r)$ at the requested initial separation. Neglecting terms of 
$O(p_{r_*}^2)$, one can then use $d p_{\varphi}/dr = \hat{\mathcal{F}}_{\varphi} \dot{r}^{-1}$ to compute 
$p_{r_*}$ at the first PA order. This procedure can then be repeated any number of times, 
with even (odd) PA orders providing corrections to $p_\varphi$ ($p_{r_*}$).
Correctly computing the initial conditions of the systems and having $p_{r_*}$ different 
from zero at the initial separation is crucial to avoid effects due to spurious eccentricity.

For eccentric binaries, initial conditions necessarily need to be specified in a 
different manner. Let us denote with $e$ the eccentricity of the ellipse that the system would orbit along assuming no GW emission. 
Similarly, let us denote with $p$ its semilatus rectum and with $\xi$ its anomaly. 
A generic point on the ellipse has radial coordinate $r = p/(1+e\cos \xi)$.
To find adiabatic initial conditions for our EOB dynamics we need to find a way to map 
$(f_0, e, \xi)$ into $(r_0, p_{\varphi}^0, p_{r_*}^0)$.
In practice, for convenience, the initial orbital frequency $\Omega_0$ is always assumed to correspond either to the apastron ($r_0 = p_0/(1-e)$), periastron ($r_0 = p_0/(1+e)$) or to the average frequency between the two points.
We then solve numerically 
\begin{equation}
\label{eq:freq0}
\partial_{p_\varphi} H(r_0(p_0), j_0(p_0), p_{r_*}=0) = \Omega_0
\end{equation}
where $j_0$ is the adiabatic angular momentum computed using energy conservation 
\begin{equation}
\hat{H}_{\rm eff}^0(p_0, j_0, \xi=0) = \hat{H}_{\rm eff}^0(p_0, j_0, \xi=\pi),
\end{equation}
and estimate the semilatus rectum of the obit $p_0$.
The evolution of the system is then always started at the apastron, 
so that 
\begin{align}
\label{eq:r0}
r_0 &= \frac{p_0}{(1-e)}, \\
\label{eq:pphi0}
p_{\varphi}^0 &= j_0,\\
\label{eq:pr*0}
p_{r_*}^0 &= 0.
\end{align}
This adiabatic procedure can be generalized to higher PA orders\footnote{1PA eccentric initial conditions have been implemented in the public \TEOBResumS{} code in commit {\tt eb5208a} }.
We leave a discussion of such initial conditions to future work.

\section{Tables}
\label{app:tables}
In this section we report the posteriors for $M_c$, $\chi_{\rm eff}$ and $q$ for two injections and different recoveries performed. 
\newline 

\begin{table*}[h!]
\begin{tabular}{c||c|c||c|c||c}
    \hline
    \hline
\multicolumn{1}{c||}{}   
&\multicolumn{5}{|c}{Circular injection}\\
\hline
Model &\TEOBResumSGIOTTO{}& \TEOBResumSecce{}&
\TEOBResumSecce{}& \TEOBResumSecce{}& \TEOBResumSecce{}\\
$e_0$-prior &$e_0 = 0$ (fixed) & $ e_0 = 10^{-8}$ (fixed) & $\mathcal{U}$(0.001,0.2) & $\mathcal{U}$(0.001,0.2) & Log-uniform(0.001, 0.2) \\
$f_0$-prior & $f_0 =$ 20 Hz (fixed) & $f_0=$ 20 Hz (fixed) & $\mathcal{U}$(18, 20.5) &$f_0 =$ 20 Hz & $\mathcal{U}$(18, 20.5) \\
\hline
$M_c (M_{\rm \odot})$ &$24.38_{-0.16}^{+0.17}$ &$24.53_{-0.17}^{+0.18}$ & $24.32_{-0.19}^{+0.18}$ & $24.33_{-0.21}^{+0.18}$ & $24.35_{-0.16}^{+0.17}$ \\ 
$\chi_{\rm eff}$ &$0.01_{-0.03}^{0.03}$ &$0.04_{-0.03}^{+0.03}$ & $0.00_{-0.03}^{+0.03}$ & $0.00_{-0.03}^{+0.03}$ & $0.00_{-0.02}^{+0.02}$ \\
$q$ &$2.00_{-0.19}^{+0.22}$ & $2.08_{-0.20}^{+0.18}$& $2.00_{-0.19}^{+0.14}$ & $2.02_{-0.18}^{+0.17}$ & $1.97_{-0.20}^{+0.17}$\\
$e_0$ & -- & -- & $0.01_{-0.01}^{+0.01}$ & $0.01_{-0.01}^{+0.01}$ & $0.00_{-0.01}^{+0.01}$ \\
\hline
\end{tabular}
\caption{\label{tab:inj_circ_posteriors} 
Posterior distribution functions for $M_c$, $\chi_{\rm eff}$ and $q$ for a circular injection ($e^{\rm inj}_{\omega} = 0$ and $f_0 = 20$Hz) with different recoveries using \TEOBResumSGIOTTO{} and \TEOBResumSecce{}.  }
\end{table*} 

\begin{table}[h!]
\begin{tabular}{c||c|c}
    \hline
    \hline
\multicolumn{1}{c||}{}   
&\multicolumn{2}{|c}{Eccentric injection}\\
\hline
Model &\TEOBResumSGIOTTO{}& \TEOBResumSecce{}\\
$e_0$-prior &$e_0 = 0$ (fixed) & $\mathcal{U}$(0.001,0.2) \\
$f_0$-prior & $f_0 =$ 20 Hz (fixed)& $\mathcal{U}$(18, 20.5)\\
\hline
$M_c (M_{\rm \odot})$ & $24.34_{-0.17}^{+0.17}$&$24.43_{-0.24}^{+0.19}$ \\ 
$\chi_{\rm eff}$ & $-0.03_{-0.03}^{+0.03}$& $-0.03_{-0.03}^{+0.03}$\\
$q$ &$1.84_{-0.21}^{+0.19}$ & $1.97_{-0.20}^{+0.17}$\\
$e_0$ &-- & $0.05_{-0.01}^{+0.01}$\\
\hline
\end{tabular}
\caption{\label{tab:inj_ecc_posteriors} Posterior distribution functions for $M_c$, $\chi_{\rm eff}$ and $q$ for an eccentric injection ($e^{\rm inj}_{\omega} = 0.05$ and $f_0 = 20$Hz) with different recoveries using \TEOBResumSGIOTTO{} and \TEOBResumSecce{}. 
}
\end{table} 

\section{Full corner plots for the GW150914 eccentric analysis}
In this section we report the full corner plots showing the posterior distributions of the intrinsic and extrinsic parameters relative to the eccentric analysis of GW150914.
\onecolumngrid
\label{app:full_corner_GW150914}
\begin{figure*}[h!]
\center
\includegraphics[width=0.98\textwidth]{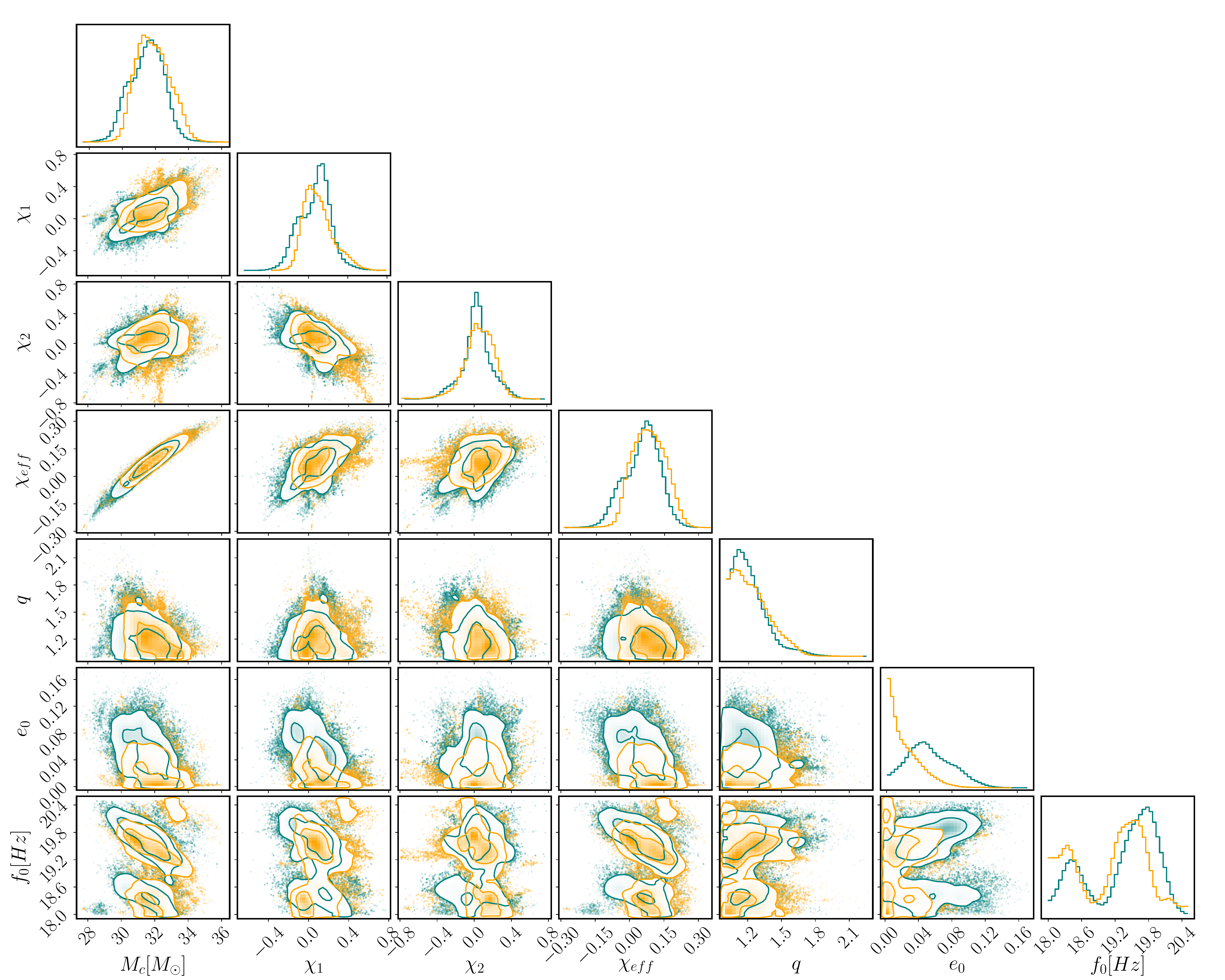}
\caption{\label{fig:corner_GW150914_intrinsic_ecc} One dimensional and join posterior distributions for the intrinsic parameters in addition with $e_0$ and $f_0$ recovered with the two eccentric analyses of GW150914. The analysis using a uniform eccentricity prior is represented in teal, the one utilizing a logarithmic-uniform prior for the eccentricity is shown in orange.}
\end{figure*}

\begin{figure*}[!tph]
\center
\includegraphics[width=0.98\textwidth]{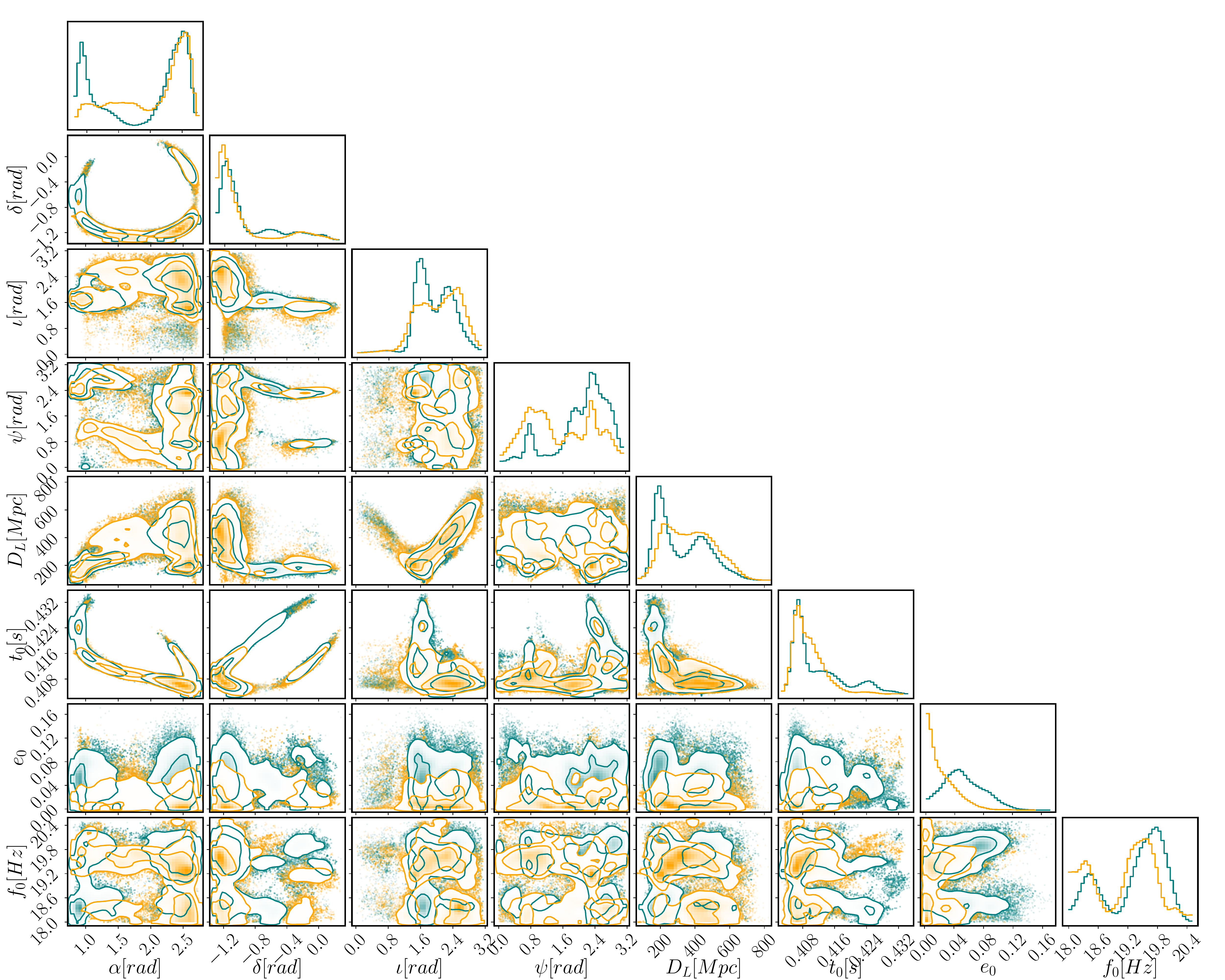}
\caption{\label{fig:corner_GW150914_extrinsic_ecc} One dimensional and join posterior distributions for the extrinsic parameters in addition with $e_0$ and $f_0$ recovered with the two eccentric analyses of GW150914. The analysis using a uniform eccentricity prior is represented in teal, the one utilizing a logarithmic-uniform prior for the eccentricity is shown in orange. }
\end{figure*}

\clearpage
\twocolumngrid
\bibliography{refs.bib}

\end{document}